\documentclass[prc,superscriptaddress,twocolumn, reprint]{revtex4}
\usepackage{hyperref}
\usepackage{mathtools}
\usepackage{amsfonts}
\usepackage{amssymb}
\usepackage{graphicx}
\usepackage{graphics}
\usepackage{listings}
\usepackage{color,soul}
\usepackage{verbatim}
\usepackage{bm}
\begin{document}

\title{Origin of hydrodynamic instability from noise: from laboratory flow to accretion disk}
\author{Subham Ghosh}
\email{subham@iisc.ac.in}
\author{Banibrata Mukhopadhyay}
\email{bm@iisc.ac.in}
\affiliation{Department of Physics, Indian Institute of 
Science, Bangalore 560012, India}

\begin{abstract}
\noindent
We attempt to address the old problem of plane shear flows: the origin of turbulence and hence transport of angular momentum in accretion 
flows as well as laboratory flows, such as plane Couette flow. We undertake the problem by introducing an 
extra force in Orr-Sommerfeld and Squire equations along with the Coriolis force mimicking the local region of the accretion disk. For 
plane Couette flow, the Coriolis term drops. Subsequently we solve the equations by WKB approximation method. We investigate the dispersion 
relation for the Keplerian flow and plane Couette flow for all possible combinations of wave vectors. Due to the very presence of extra 
force, we show that both the flows are unstable for a certain range of wave vectors. However, the nature of instability between the flows 
is different. We also study the Argand diagrams of the perturbation eigenmodes. It helps us to compare the different time scales 
corresponding to the perturbations as well as accretion. We ultimately conclude with this formalism that fluid gets enough time to be 
unstable and hence plausibly turbulent particularly in the local regime of the Keplerian accretion disks. Repetition of the analysis 
throughout the disk explains the transport of angular momentum and matter along outward and inward direction respectively.  
\end{abstract}


\maketitle

\section{Introduction}


A long-standing mismatch between theory and experiment regarding the transition from laminar to turbulent flows for laboratory fluids, e.g. 
plane Couette flow and plane Poiseuille flow, is there in literatures. The linear theory of perturbation says that plane 
Poiseuille flow becomes unstable beyond Reynolds number ($Re$) 5772.22 \cite{Orszag_1971} and, on the other hand, plane Couette flow is 
stable for any $Re$ \cite{Romanov_1973}. However, according to experiments/simulations, beyond $Re \sim 1000$ \cite{Alavyoon_1986, 
Grossmann_2000} and $Re \sim 350$ \cite{Dauchot_1994, Duguet_2010} the laminar flow becomes turbulent in case of plane Poiseuille flow and 
plane Couette flow respectively. The similar kind of mismatch is there in the context of astrophysics particularly in case of accretion 
disks. Accretion disks are astrophysical objects formed around a denser object mainly in the form of a disk. Nevertheless, the accretion 
disk 
involves very sophisticated (or rich) physics behind the formation and evolution of its various parts depending on the nature of the 
central objects (black holes, white dwarfs, neutron stars, main sequence stars, etc) around which the matter accretes in the form of a 
disk. The physics also involves with the nature of mass supply (e.g. mass supplied from evolved stars, from interstellar medium, 
molecular cloud, etc) that helps accretion around the central object. However, in this work, we shall be discussing a geometrically thin 
and optically thick disk, where the accreting matter almost follows Kepler's law, i.e. the fluid particle in the corresponding flow 
revolves around the central object at a particular radius due to the almost balance between inward gravitational force and outward 
centrifugal force. The flow therefore is called Keplerian flow. The change in the angular momentum per unit mass of the fluid particle, 
therefore, occurs in increasing proportion to the square root of the radial distance of the particle. Due to the very nature of the Keplerian 
rotation, the perturbation of the fluid particle decays down and eventually the particle returns to its initial position. This is called 
Rayleigh stability. The Keplerian flow, therefore, is Rayleigh stable.

However, due to the Keplerian rotation, two fluid layers across the radial direction in the disk will have different angular velocities. 
Since the flow has differential velocity across the radial direction, molecular viscosity comes into picture. However, observational 
evidences, e.g. temperature, luminosity, etc. from the Keplerian accretion disk do not support the molecular viscosity as the origin of 
matter transport. The molecular viscosity is so weak that it cannot transport the angular momentum outward and matter inward and hence 
cannot explain the observables \cite{Frank_2002}.  The belief is that it is the turbulent viscosity which is behind the transport. The idea 
was put forward by Shakura \& Sunyaev \cite{Shakura_1972}  and  Lynden-Bell \& Pringle \cite{Lynden-Bell_1974} without 
explicitly revealing the reason behind the turbulence. In 1991, Balbus \& Hawley \cite{Balbus_1991} came up with an idea of instability 
mechanism due to the interplay between weak magnetic field and the rotation of the fluid parcel, naming magneto-rotational instability 
(MRI), following the idea of Velikhov \cite{Velikhov_1959} and Chandrasekhar \cite{Chandrasekhar_1960}. In spite of the 
overwhelming success of MRI in explaining the origin of turbulence, it is not 
out of caveats. In the colder systems, e.g. protoplanetary 
disk \cite{Bai_2013_ApJ, Bai_2017}, cataclysmic variables in their low states \cite{Gammie_1998, Menou_2000}, the outer 
part of active 
galactic nucleus (AGN) disks and the underlying dead zone \cite{Menou_2001}, where the ionization is very small such that 
matter cannot 
be coupled with the magnetic field, MRI gets suppressed. It is not only the low ionization that challenges MRI, there are, in fact, a lot 
of other examples too. Nath $\&$ Mukhopadhyay \cite{Nath_2015} argued that it is the magnetic transient growth that brings 
nonlinearity and hence plausible turbulence in the system beyond Reynolds number ($Re$) $10^9$, since their growth rate is faster than MRI 
in that regime. Usually, $Re$ in accretion disks \cite{Mukho_2013} is larger than this value, hence the relevance of MRI 
in large $Re$ 
systems is questionable. As a general interest, the transient energy growth in the case of magnetohydrodynamical shear flows (with 
viscosity and resistivity included) was studied further by Bhatia \& Mukhopadhyay \cite{bhatia_2016}. They showed that even transient 
energy growth ceased to occur beyond certain magnetic field. In addition to this, Pessah $\&$ Psaltis \cite{Pessah_2005} and Das et al. 
\cite{Das_2018}, using local and global analysis respectively, showed the stabilization of the axisymmetric MRI above a certain magnitude of 
a toroidal component of the magnetic field for compressible and differentially rotating flows. It is, therefore, of great concern whether 
there is any instability in the system from hydrodynamical origin. 

However, in the literature \cite{Dubrulle_2005_a, Dubrulle_2005_b, Dauchot_1995, Rudiger_2001, Klahr_2003, Richard_1999, Kim_2000, 
Mahajan_2008, Yecko_2004, Lesur_2010, Mukhopadhyay_Mathew_2011, Mukhopadhyay_2013}, there is a long standing 
debate regarding the stability of Rayleigh 
stable flows, particularly in the context of accretion disks. Approximating the local hot accretion flow to be shearing sheet, 
people \cite{Balbus_1996, Hawley_1999} attempted, analytically and with simulation, to resolve the issue without 
considering viscosity. 
They concluded that the sustained turbulence and hence outward transport of angular momentum were not 
possible in the Keplerian flow if hydrodynamics was considered only. However, Lesur \& Longaretti \cite{Lesur_2005} with 
shearing sheet approximation and considering viscosity strongly disagreed with the aforementioned authors and claimed that 
the absence of turbulence in the simulation in the above mentioned works was resolution issue. Although they agreed 
that there was lack of computer resources to resolve the Keplerian regime, their extrapolated numerical data could not produce
astrophysically sufficient subcritical turbulent transport in the Keplerian flow. Pumir \cite{Pumir_1996} claimed for 
sustained turbulence if the mean flow is plane Couette typed. However, they did not consider rotational effects. Fromang 
and Papaloizou \cite{Fromang_2007}, though did magnetohydrodynamical (MHD) simulation, argued for
considering explicit diffusion coefficients: both resistive and viscous, whose effect is stronger than numerical 
dissipation effect, before making any conclusion based on MHD simulation. Therefore, we notice that in 
all of these works some important physics are missing, i.e., viscosity \cite{Balbus_1996, Hawley_1999}, resolution of 
the Keplerian region \cite{Lesur_2005}, the Coriolis force \cite{Pumir_1996}, explicit diffusion coefficients (both 
viscous and resistive) \cite{Fromang_2007} are not adequately considered. Even if we have well-resolved simulations 
\cite{Nauman_2016, 
Shi_2015, Walker_2016}, the previously mentioned facts or parameter regions exist, where MRI is 
inapplicable/insufficient as an instability mechanism. Nevertheless, the authors argued for plausible 
emergence of hydrodynamics instability and hence further turbulence by experiment (e.g. \cite{Paoletti_2012}), simulations in the
context accretion disks (e.g. \cite{Avila_2012}), transient growth in the case of otherwise linearly stable flows (e.g. \cite{man_2005, 
amn_2005, Mukhopadhyay_Mathew_2011, Cantwell_2010}). 

We, therefore, search for a hydrodynamical origin of nonlinearity and hence plausible turbulence in the accretion disk. We, in particular, 
consider an extra force in this work and the force has stochastic origin. The existence and consequences of the stochastic force in the 
hydrodynamical systems were initiated by Mukhopadhyay \& Chattopadhyay \cite{Mukhopadhyay_2013} inspired by the idea of Nelson \& 
Foster \cite{Forster_1977} and DeDominicis \& Martin \cite{DeDominicis_1979}. They showed that the presence of the stochastic 
force in the rotating shear flows in a narrow gap limit reveals large correlation of energy growth of the perturbation. Later, Nath \& 
Mukhopadhyay \cite{Nath_2016} obtained the dispersion relation of the linear perturbations considering stochastic force in the 
Orr-Sommerfeld and Squire equations, describing the fluid flow in a small radial patch of accretion disk. However, they considered plane 
wave perturbation with constant amplitude as the trial solution of the Orr-Sommerfeld and Squire equations. In the present work, we 
consider three-dimensional perturbations and WKB approximation to obtain the solutions for Orr-Sommerfeld and Squire equations. 
While qualitatively we obtain similar result as Nath \& Mukhopadhyay \cite{Nath_2016}, it brings new quantitative insight which is useful to infer
observed data and/or experimental results based on our model. We also obtain the Argand diagrams 
corresponding to the perturbations and these are necessary to compare the timescales corresponding to the growth with that of oscillation of the 
perturbations. In addition to this, we also confirm whether the fluid parcel inside the shearing box within a small patch of accretion disk 
gets enough time to enter into the nonlinear regime and hence becomes turbulent within the timescale it came across the box. However, for 
plane Couette flow, we do not need to worry about any such time scale, as there is no radial infall.

The plan of the paper is the following. In \S \ref{sec:Formalism}, we describe the governing equations which are Orr-Sommerfeld and 
Squire equations in the presence of Coriolis force and noise for linearly perturbed flow inside a shearing box at a smaller patch of 
accretion disk. We then write them in the Fourier space to obtain a general dispersion relation. In \S \ref{sec:dis_rel}, the dispersion 
relation is studied extensively for the Keplerian and plane Couette flows. The Argand diagrams corresponding to the linear perturbations 
in the case of Keplerian flow are studied in \S \ref{sec:arg_dia} for various parameters. In the end, we discuss about the plausibility of 
occurrence of instability which could further lead to nonlinearity and hence turbulence in the context of accretion disks and laboratory 
flows, e.g. plane Couette flows in \S \ref{sec:discusion}. We finally conclude in \ref{sec:Conclusion} that our model is 
able to explain the origin of instability and hence turbulence in the context of accretion disk as well as plane Couette 
flow.

\section{Formalism}
\label{sec:Formalism}
The detailed description of the local formulation can 
be found in Mukhopadhyay et al. \cite{man_2005} and also in Bhatia \& Mukhopadhyay \cite{bhatia_2016}. The schematic diagram of 
the background flow inside the shearing box is shown in \cite{Mukhopadhyay_2011NJPh}. As the fluid is in the local region, we assume the 
fluid to be incompressible 
\cite{amn_2005, Nath_2015}. There we recast the Navier-Stokes equation in Orr-Sommerfeld and Squire equations in the 
presence of Coriolis 
force and extra force, by eliminating the pressure term from different components of the Navier-Stokes equation and utilizing the 
continuity equation for incompressible flow \cite{Nath_2016}. The ensemble averaged Orr-Sommerfeld and Squire equations 
in the presence 
of Coriolis force and extra force are given by
\begin{equation}
   \left(\frac{\partial}{\partial t}+U\frac{\partial}{\partial y}\right)\nabla^2u -\frac{\partial^2 U}{\partial x^2} \frac{\partial 
u}{\partial y}+ \frac{2}{q}\frac{\partial \zeta}{\partial z} = \frac{1}{Re}\nabla^4u +\eta_1,
	\label{eq:gen_velo_pertb}
\end{equation}

\begin{equation}
   \left(\frac{\partial}{\partial t}+U\frac{\partial}{\partial y}\right)\zeta -\left(\frac{\partial U}{\partial x} + \frac{2}{q}\right) 
\frac{\partial u}{\partial z} = \frac{1}{Re}\nabla^2 \zeta +\eta_2,
	\label{eq:gen_vorti_pertb}
\end{equation}
where $U = -x$ is the $y$ -component of background velocity. The other components of background velocity are zero; 
$u$ and $\zeta$ are $x$ -components of velocity and vorticity perturbations respectively; $q$ is the rotation parameter which 
describes the radial dependence of the angular frequency of fluid element around the central object, given by $\Omega 
\propto 1/r^q$; $Re$ is the Reynolds number; $\eta_1$ and $\eta_2$ are the extra forces on the fluid particles.
$q$ becomes 1.5 and $\infty$ for the Keplerian and plane Couette flows \cite{man_2005, bhatia_2016} respectively. 
In order to obtain the dispersion relation, we write down the above equations in the Fourier space. Our conventions for
Fourier transform  and inverse Fourier transform are respectively
\begin{equation}
 A(\textbf{r},t) = \int{\tilde{A}_{\textbf{k},\omega}}e^{i(\textbf{k}\cdot \textbf{r} - \omega t)}d^3k\ d\omega
 \label{eq:Fourier_trans}
\end{equation} 
and
\begin{equation}
 \tilde{A}_{\textbf{k},\omega} = \left(\frac{1}{2\pi}\right)^4\int{A(\textbf{r},t)e^{-i(\textbf{k}\cdot \textbf{r} - \omega t)}} d^3x\ dt.
 \label{eq:inv_Fourier_trans}
\end{equation}
Here $A$ can be any one of $u, \zeta,$ and $\eta_i;$ $\textbf{k}$ and $\omega$ are the wavevector and frequency, respectively, in Fourier 
space such that in Cartesian coordinates $\textbf{k}=(k_x,k_y,k_z)$ and $|\textbf{k}| = k;$ $\textbf{r}$ is the position vector and in 
Cartesian coordinates $\textbf{r} = (x,y,z).$

The boundary conditions to solve equations (\ref{eq:gen_velo_pertb}) and (\ref{eq:gen_vorti_pertb}) are
\begin{equation}
 u=\frac{\partial{u}}{\partial{x}} = \zeta = 0, \ {\rm at}\ x = \pm 1. 
 \label{eq:boundary_condition} 
\end{equation}
\\
In Fourier space, equations (\ref{eq:gen_velo_pertb}) and (\ref{eq:gen_vorti_pertb}) become 
 \begin{eqnarray}
\begin{split}
 k_yk^2 \frac{\partial  \tilde{u}_{\textbf{k},\omega}}{\partial k_x} =  \left(i\omega k^2 - 2k_xk_y - 
\frac{k^4}{Re}\right) 
\tilde{u}_{\textbf{k},\omega} 
 \\ +\frac{2ik_z}{q} \tilde{\zeta}_{\textbf{k},\omega} -\ m_1\delta(\textbf{k})\delta(\omega),
\end{split}
\label{eq:Keplerian_velo_Fourier_space}
\end{eqnarray}
\begin{eqnarray}
 \begin{split}
  k_y \frac{\partial  \tilde{\zeta}_{\textbf{k},\omega}}{\partial k_x} =  -ik_z \left(1-\frac{2}{q}\right) 
\tilde{u}_{\textbf{k},\omega}
   +\left(i\omega - \frac{k^2}{Re}\right) \tilde{\zeta}_{\textbf{k},\omega} \\+\ m_2\delta(\textbf{k})\delta(\omega),
 \end{split}
 \label{eq:Keplerian_vorti_Fourier_space}
\end{eqnarray}
where the Fourier transform of $\eta_i$ is $m_i\delta(\textbf{k})\delta(\omega)$ with $m_i$ being the constant mean corresponding to 
$\eta_i$. 
The traveling wave solutions for equations 
(\ref{eq:gen_velo_pertb}) and (\ref{eq:gen_vorti_pertb}) are assumed to be 
\begin{equation}
u = u(x) e^{i(\bm{\alpha} \cdot \textbf{r} - \beta t)},
\zeta = \zeta(x) e^{i(\bm{\alpha} \cdot \textbf{r} - \beta t)},
\label{eq:lin_sol}
\end{equation}
where the wave vector, $\bm{\alpha}$, is given by $\bm{\alpha} = (\alpha_1, \alpha_2, \alpha_3)$, and $\beta$ is the 
frequency. Usually $\beta$ is a complex quantity and, according to our convention, if the 
imaginary part of $\beta$, i.e. $Im(\beta)$, is positive, then the perturbation grows with time. To obtain the dispersion relation, we 
transform equation (\ref{eq:lin_sol}) in the Fourier space (see Appendix \ref{sec:obtaining_des_rel}) and substitute them in the equations 
(\ref{eq:Keplerian_velo_Fourier_space}) and 
(\ref{eq:Keplerian_vorti_Fourier_space}) and then we integrate with respect to $\omega$ and $k$. See Appendix \ref{sec:obtaining_des_rel}
for details. We further use WKB approximation to obtain the solution. Therefore, we neglect second and higher order derivatives, as they are varying slowly over 
the length $1/\alpha_1$. The dispersion relations from equations (\ref{eq:Keplerian_velo_Fourier_space}) and 
(\ref{eq:Keplerian_vorti_Fourier_space}) are then
\begin{eqnarray}
 \begin{split}
    \left(i\beta\alpha^2 - \frac{\alpha^4}{Re}\right)u(0)+2i\alpha_1\left(\frac{2\alpha^2}{Re} - 
i\beta\right)u'(0)\\+\frac{2i\alpha_3}{q}\zeta(0) - m_1 = 0\\
-i\alpha_3 \left(1-\frac{2}{q}\right)u(0)+\left(i\beta-\frac{\alpha^2}{Re}\right)\zeta(0)\\+\frac{2i\alpha_1}{Re}\zeta'(0)+m_2 = 0.
 \label{eq:dispersion_rel_orr_som_squ}
 \end{split}
\end{eqnarray}
Here, $u(0)$ and $u'(0)$ are respectively values of $u(x)$ and $u'(x)$ at $x=0$.
We also consider the first order derivatives to be
\begin{eqnarray*}
	  \begin{array}{ll}
	  u'(0) = \gamma u(0) = \gamma u_0, \\
	  \zeta'(0) = \gamma \zeta(0) = \gamma \zeta_0, \\
	  \end{array}
\end{eqnarray*}
and the same strength for the extra forces, i.e. $m_1 = m_2 = m$. Now if we eliminate $\zeta$ with all the assumptions from equations 
(\ref{eq:Keplerian_velo_Fourier_space}) and (\ref{eq:Keplerian_vorti_Fourier_space}), we obtain the dispersion relation, which is given by 

\begin{equation}
\begin{split}
 m \left(2 \alpha _3+\beta  q+\frac{i \alpha ^2 q}{Re}+\frac{2 \alpha _1 \gamma  q}{Re}\right)=u_0 \Bigl(2 i \alpha _3^2\\+i 
\alpha ^2 \beta ^2 q+2 \alpha _1 \beta ^2 \gamma  q-\frac{4 i \alpha _3^2}{q}-\frac{i \alpha ^6 q}{Re^2}-\frac{6 \alpha _1 \alpha ^4 
\gamma  q}{Re^2}\\+\frac{8 i \alpha _1^2 \alpha ^2 \gamma ^2 q}{Re^2}-\frac{2 \alpha ^4 \beta  q}{Re}+\frac{8 i \alpha 
_1 
\alpha ^2 \beta  \gamma  q}{Re}+\frac{4 \alpha _1^2 \beta  \gamma ^2 q}{Re}\Bigr).
\end{split}
\label{eq:dispersion_rel}
\end{equation} 
For clarity, we consider $\gamma = \pm 1, \pm \alpha_1, \pm i\alpha_1.$ However, only $\gamma = i\alpha_1$ gives $Im(\beta)<0$ for any $Re$ 
without 
extra force is considered, i.e. $m = 0$, which is physical. We, therefore, stick to $\gamma = i\alpha_1$ throughout the paper.
For the computational purpose, we consider the components of wave vectors along $y$-direction 
to be zero, i.e. $\alpha^2 = \alpha_1^2+\alpha_3^2$. However, if 
we make $\alpha_3 = 0$ and $\alpha^2 = \alpha_1^2+\alpha_2^2$, from equation (\ref{eq:dispersion_rel_orr_som_squ}) it is clear that the 
problem will become qualitatively plane Couette flow.

\section{Dispersion Relation}
\label{sec:dis_rel}
\subsection{Keplerian flow}
\label{subsec:kep_flow}
\begin{figure}
\includegraphics[width = \columnwidth]{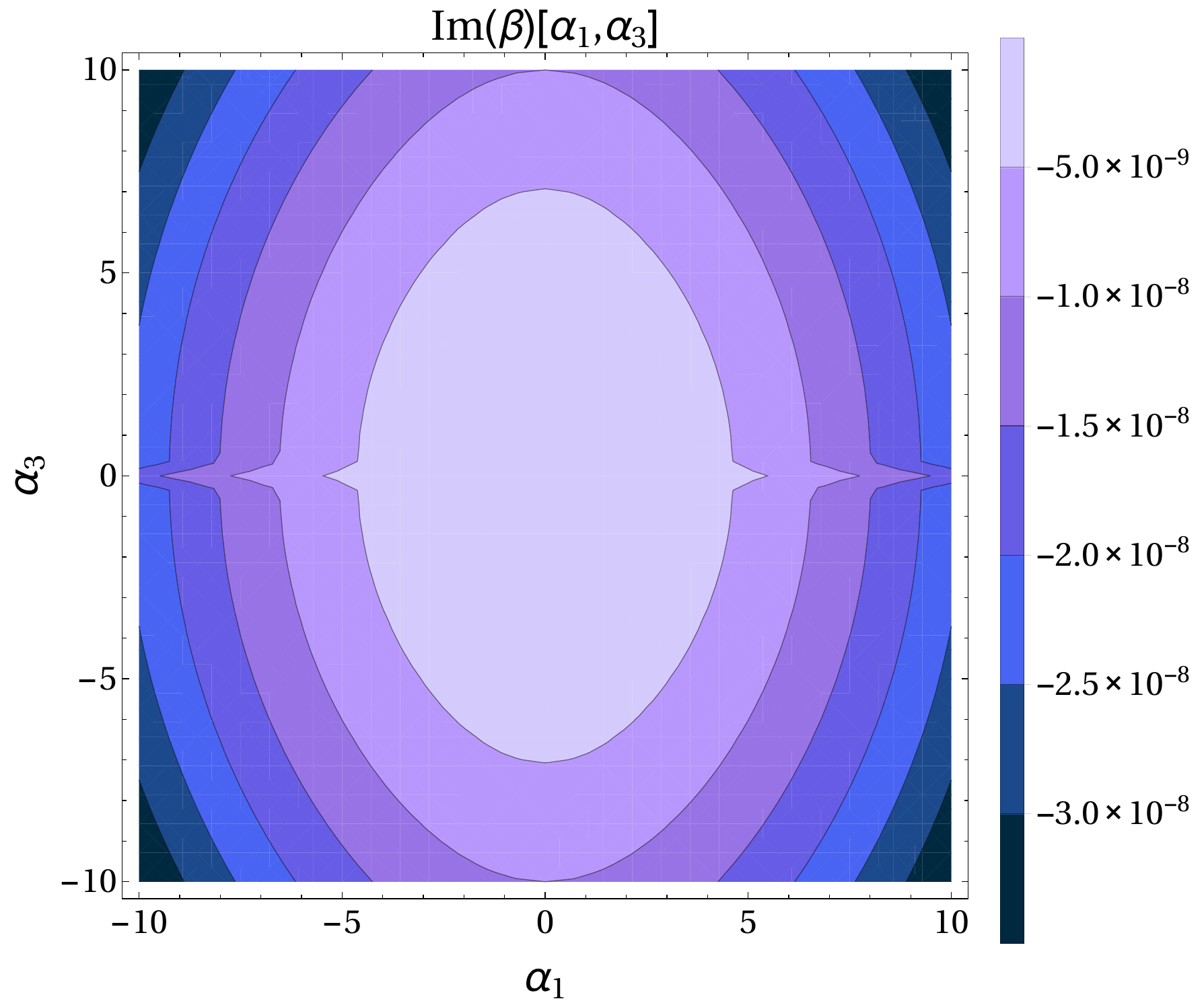}
\caption{\label{fig:keplarian_wkb_v_0_re_10_10} Variation of $Im(\beta)$ as a function of $\alpha_1$ and $\alpha_3$ for $Re = 10^{10}$ and 
$m/u_0 = 0$ for 
the Keplerian flow.}
\end{figure}

\begin{figure}
 \includegraphics[width=\columnwidth]{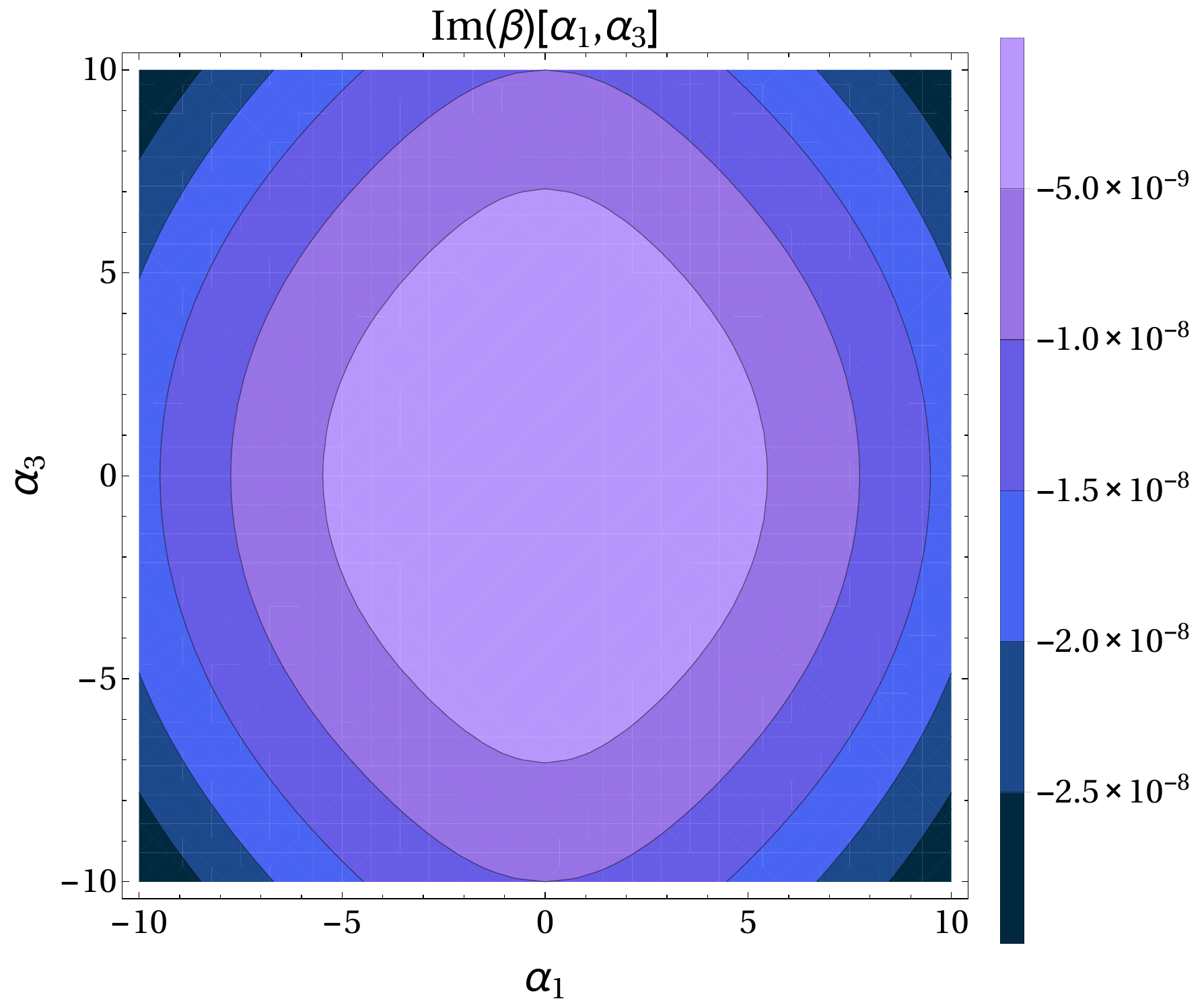}
  \caption{\label{fig:plane_cou_wkb_v_0_re_10_10} Variation of $Im(\beta)$ as a function of $\alpha_1$ and $\alpha_3$ for $Re = 10^{10}$ 
and $m/u_0 = 0$ for 
plane Couette flow.}
\end{figure}

Here we shall study the solutions of equation (\ref{eq:dispersion_rel}) for different parameters. Equation (\ref{eq:dispersion_rel}) is a 
quadratic equation of $\beta$ with complex coefficients. Among the two solutions of $\beta$, the one which we are interested in is 
\begin{eqnarray}
\begin{split}
\beta = -\frac{0.5 i}{3 \alpha _1^2 q+\alpha _3^2 q}\Bigl[\frac{m q}{u_0}+\frac{14 \alpha _1^4 q}{Re}+\frac{12 \alpha _3^2 \alpha _1^2 
q}{Re} & \\ +\frac{2 \alpha _3^4 q}{Re} -\frac{1}{Re}\Bigl(\frac{24i \alpha _3 \alpha _1^2 m q Re^2}{u_0} +\frac{8i \alpha _3^3 
m q Re^2}{u_0} &\\+\frac{m^2 q^2 Re^2}{u_0^2} -\frac{8 \alpha _1^4 m q^2 Re}{u_0}+16\alpha _1^8 q^2 +24\alpha _3^2 \alpha _1^2 q Re^2 &\\+8 
\alpha _3^4 q Re^2 -48 \alpha _3^2 \alpha _1^2 Re^2-16\alpha _3^4 Re^2\Bigr)^{\frac{1}{2}}\Bigr].
\end{split}
\label{eq:sol_beta}
\end{eqnarray}
The other solution of $\beta$ is always stable irrespective of extra force.
However, equation (\ref{eq:sol_beta}) expectedly provides negative $Im(\beta)$ for $m = 0$ irrespective of
$Re$. Interestingly, equation (\ref{eq:sol_beta}) also provides positive $Im(\beta)$ within a particular window of 
$\alpha_1$ and $\alpha_3$ beyond certain $m$ depending on $Re$ for a fixed $q$. Here we observe the 
dispersion relations, i.e. the variation of $Im(\beta)$ as a function of 
$\alpha_1$ and $\alpha_3$ for different $Re$ and $m/u_0$ for the Keplerian flow. 
FIGs.~\ref{fig:keplarian_wkb_v_0_re_10_10} and \ref{fig:plane_cou_wkb_v_0_re_10_10} show the variation of $Im(\beta)$ as a 
function of $\alpha_1$ and $\alpha_3$ for the Keplerian and plane Couette flow (see \ref{subsec:plane_cou_flow}) 
respectively for $m/u_0 = 0.$ From linear stability analysis, we know these two flows are stable for any $Re$ and this is 
confirmed in the FIGs.~\ref{fig:keplarian_wkb_v_0_re_10_10} and \ref{fig:plane_cou_wkb_v_0_re_10_10}. The kinks in 
FIG.~\ref{fig:keplarian_wkb_v_0_re_10_10} around $\alpha_3 = 0$ are there for $q<2$ and hence their presence is due to the 
rotation in the system.

The color codes that we use for the contour plots for 
FIG.~\ref{fig:keplarian_wkb_v_0_re_10_10} to FIG.~\ref{fig:kep_wkb_v_10_2_re_10_4} are the following. We use bluish 
and reddish colors to indicate $Im(\beta)$'s negativity and positivity respectively. We further use white color to 
indicate the transition from the negative to positive of $Im(\beta)$.

\begin{figure}
 \includegraphics[width=\columnwidth]{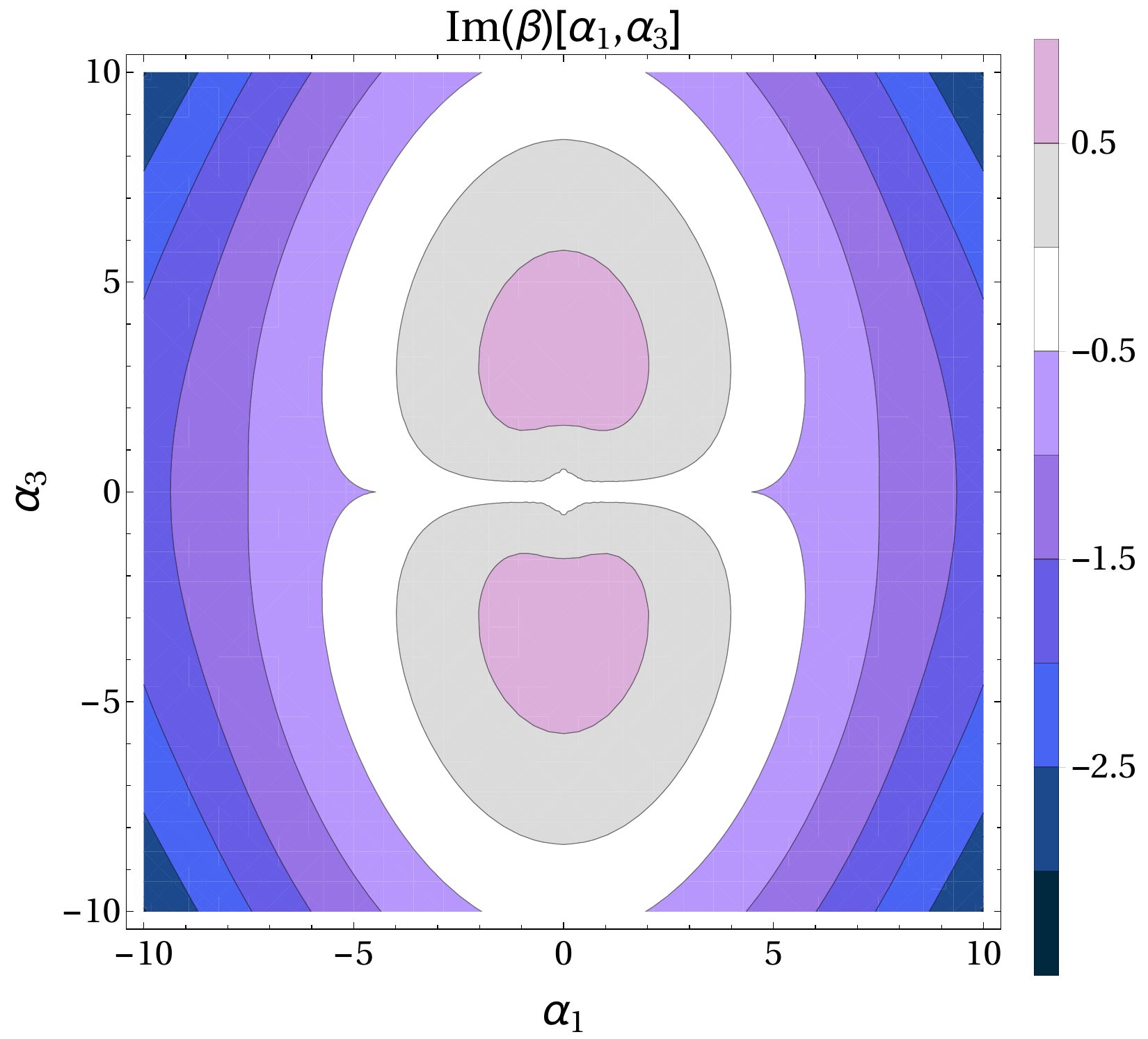}
  \caption{Variation of $Im(\beta)$ as a function of $\alpha_1$ and $\alpha_3$ for $Re = 10^{2}$ and $m/u_0 = 10$ for 
the Keplerian flow.}
\label{fig:keplarian_wkb_v_10_re_10_2}
\end{figure}

\begin{figure}
 \includegraphics[width=\columnwidth]{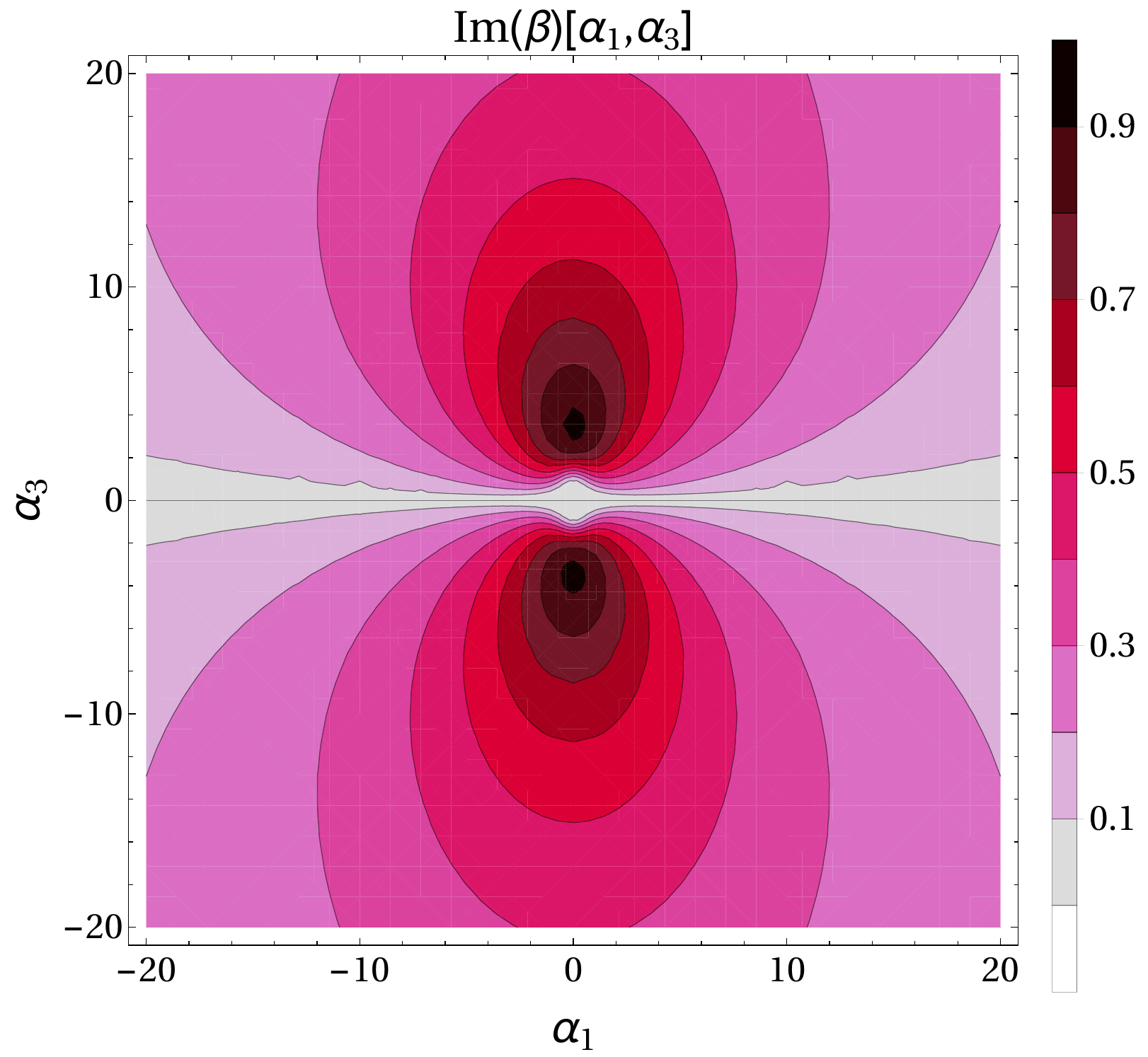}
  \caption{Variation of $Im(\beta)$ as a function of $\alpha_1$ and $\alpha_3$ for $Re = 10^{10}$ and $m/u_0 = 10$ for 
the Keplerian flow.}
\label{fig:keplarian_wkb_v_10_re_10_10}
\end{figure}

\begin{figure}
 \includegraphics[width=\columnwidth]{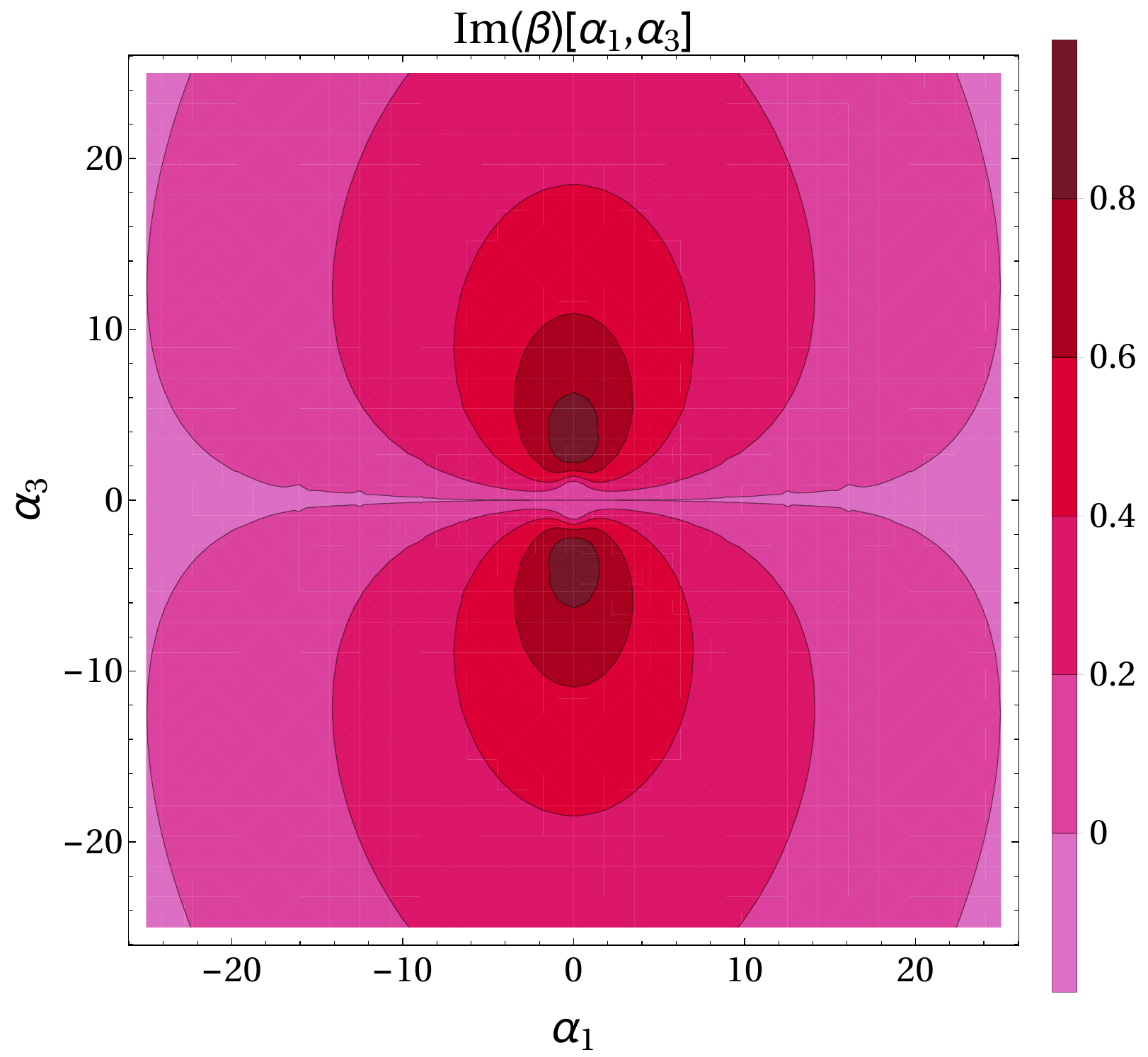}
  \caption{Variation of $Im(\beta)$ as a function of $\alpha_1$ and $\alpha_3$ for $Re = 10^{4}$ and $m/u_0 = 10$ for 
the Keplerian flow.}
\label{fig:kep_wkb_v_10_re_10_4}
\end{figure}

\begin{figure}
 \includegraphics[width=\columnwidth]{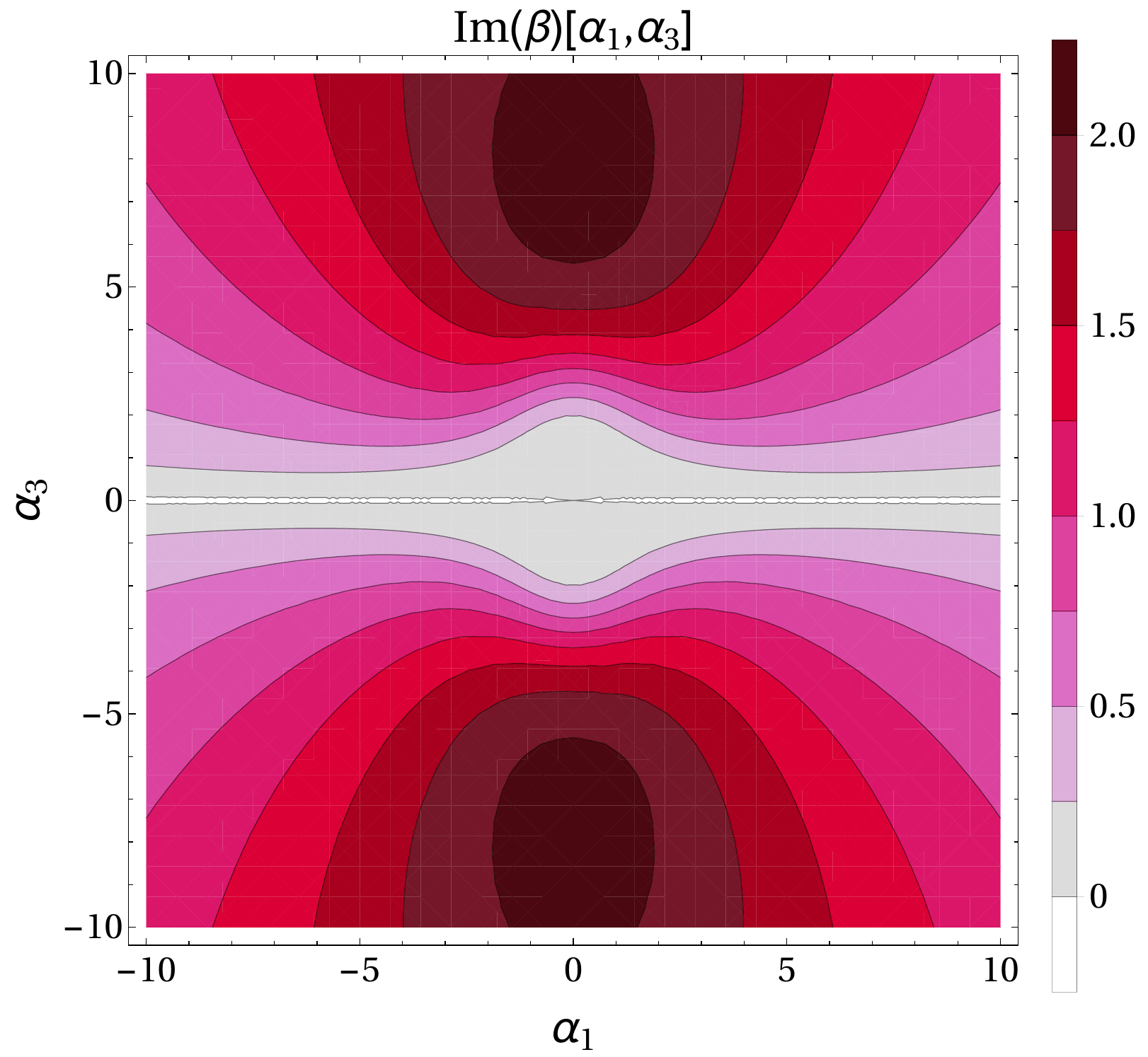}
  \caption{Variation of $Im(\beta)$ as a function of $\alpha_1$ and $\alpha_3$ for $Re = 10^{4}$ and $m/u_0 = 10^2$ for 
the Keplerian flow.}
\label{fig:kep_wkb_v_10_2_re_10_4}
\end{figure}

As we introduce the extra force, i.e. $m\neq0$, $Im(\beta)$ becomes positive for a particular range of $\alpha_1$ and 
$\alpha_3$. Throughout the paper, we use $Im(\beta)_{max}$ and $\mathcal{R}e(\beta)_{max}$ to indicate the maximum value 
of $Im(\beta)$ and at which $\mathcal{R}e(\beta)$, it occurs, respectively. FIGs.~\ref{fig:keplarian_wkb_v_10_re_10_2} 
and \ref{fig:keplarian_wkb_v_10_re_10_10} show the variation of $Im(\beta)$ as a function of $\alpha_1$ and $\alpha_3$ 
for $m/u_0 = 10$ for the Keplerian flow but for $Re = 10^2$ and $10^{10}$ 
respectively.FIGs.~\ref{fig:kep_wkb_v_10_re_10_4} and \ref{fig:kep_wkb_v_10_2_re_10_4} show the variation of $Im(\beta)$ 
as a function of $\alpha_1$ and $\alpha_3$ for $Re = 10^4$ for the Keplerian flow but for $m/u_0 = 10$ and $10^2$ 
respectively. These two figures depict that the increment of $m/u_0$ increases $Im(\beta)$ value for a fixed $Re$. 
Note that the bounds on the axes of FIGs.~\ref{fig:keplarian_wkb_v_10_re_10_10} and \ref{fig:kep_wkb_v_10_re_10_4} are 
different than that of FIGs.~\ref{fig:keplarian_wkb_v_10_re_10_2} and \ref{fig:kep_wkb_v_10_2_re_10_4}. The reason is 
described later in this section itself.
  
\begin{figure}
 \includegraphics[width=\columnwidth]{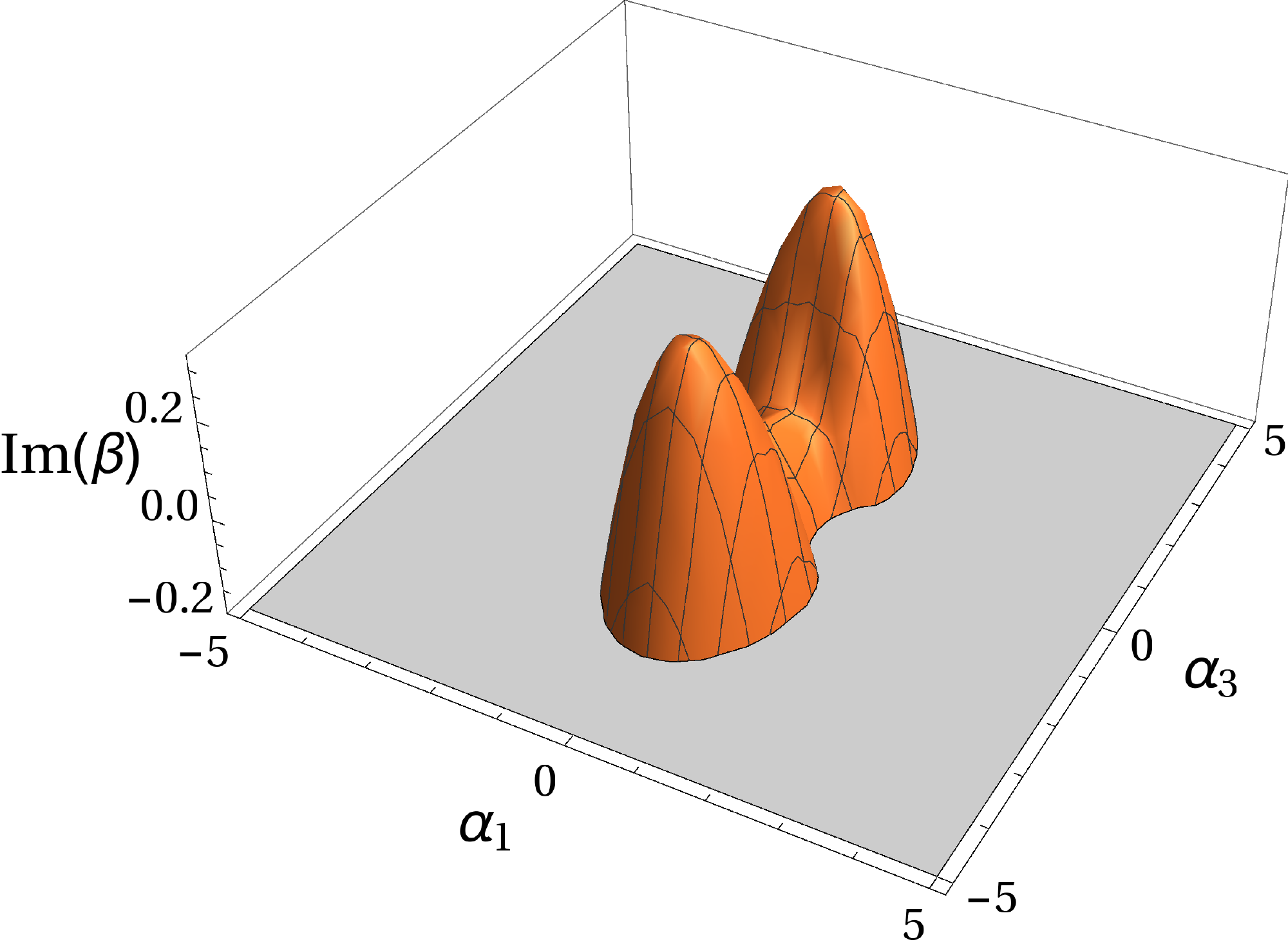}
  \caption{Variation of $Im(\beta)$ in three dimensions as a function of $\alpha_1$ and $\alpha_3$ for $Re = 10$ and $m/u_0 = 10$ for 
the Keplerian flow. $Im(\beta)_{max} = 0.33$.}
\label{fig:kep_wkb_3d_plot_v_10_re_10}
\end{figure}

\begin{figure}
 \includegraphics[width=\columnwidth]{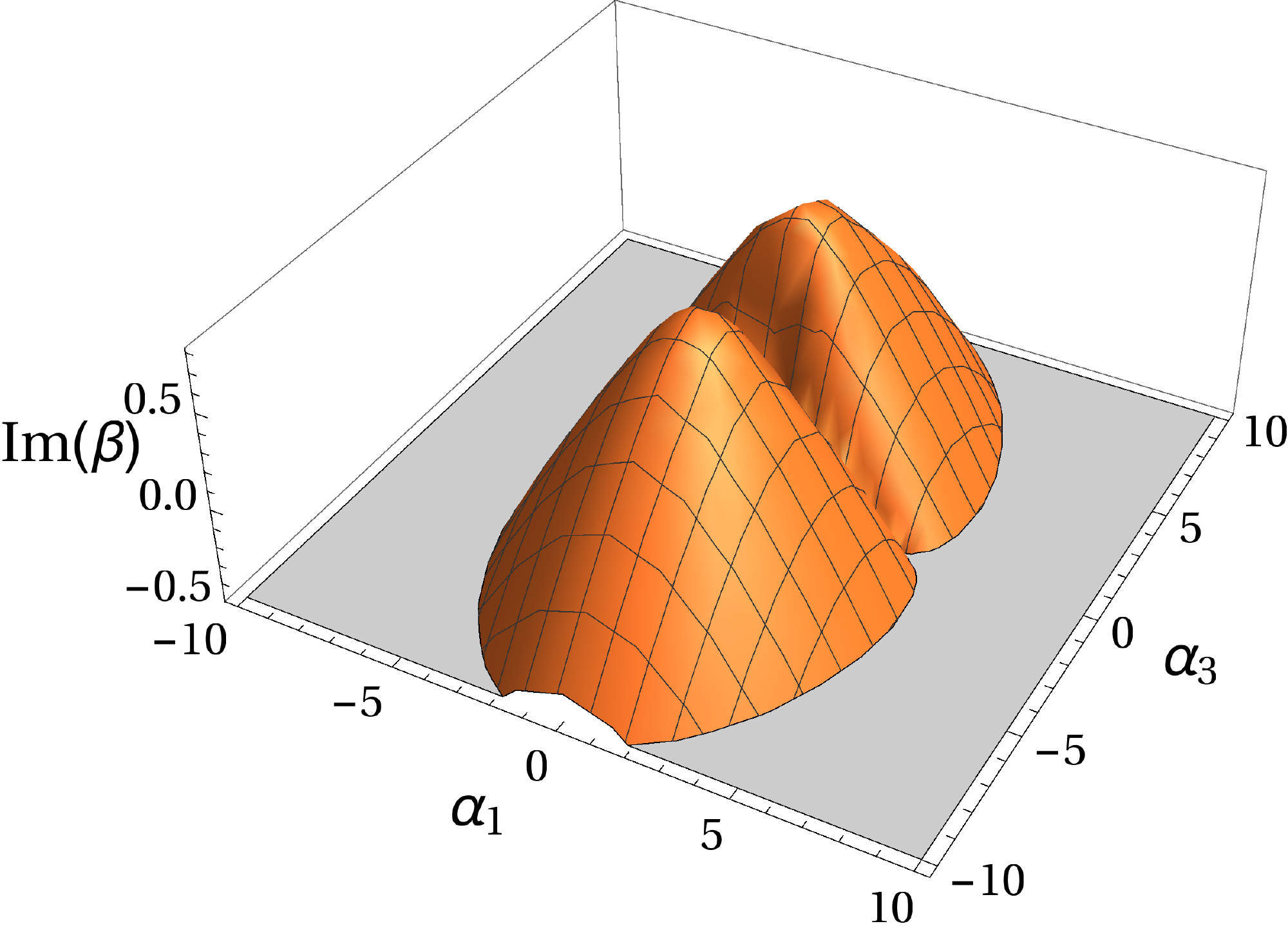}
  \caption{Variation of $Im(\beta)$ in three dimensions as a function of $\alpha_1$ and $\alpha_3$ for $Re = 10^2$ and $m/u_0 = 10$ for 
the Keplerian flow. $Im(\beta)_{max} = 0.822$.}
\label{fig:kep_wkb_3d_plot_v_10_re_10_2}
\end{figure}

\begin{figure}
 \includegraphics[width=\columnwidth]{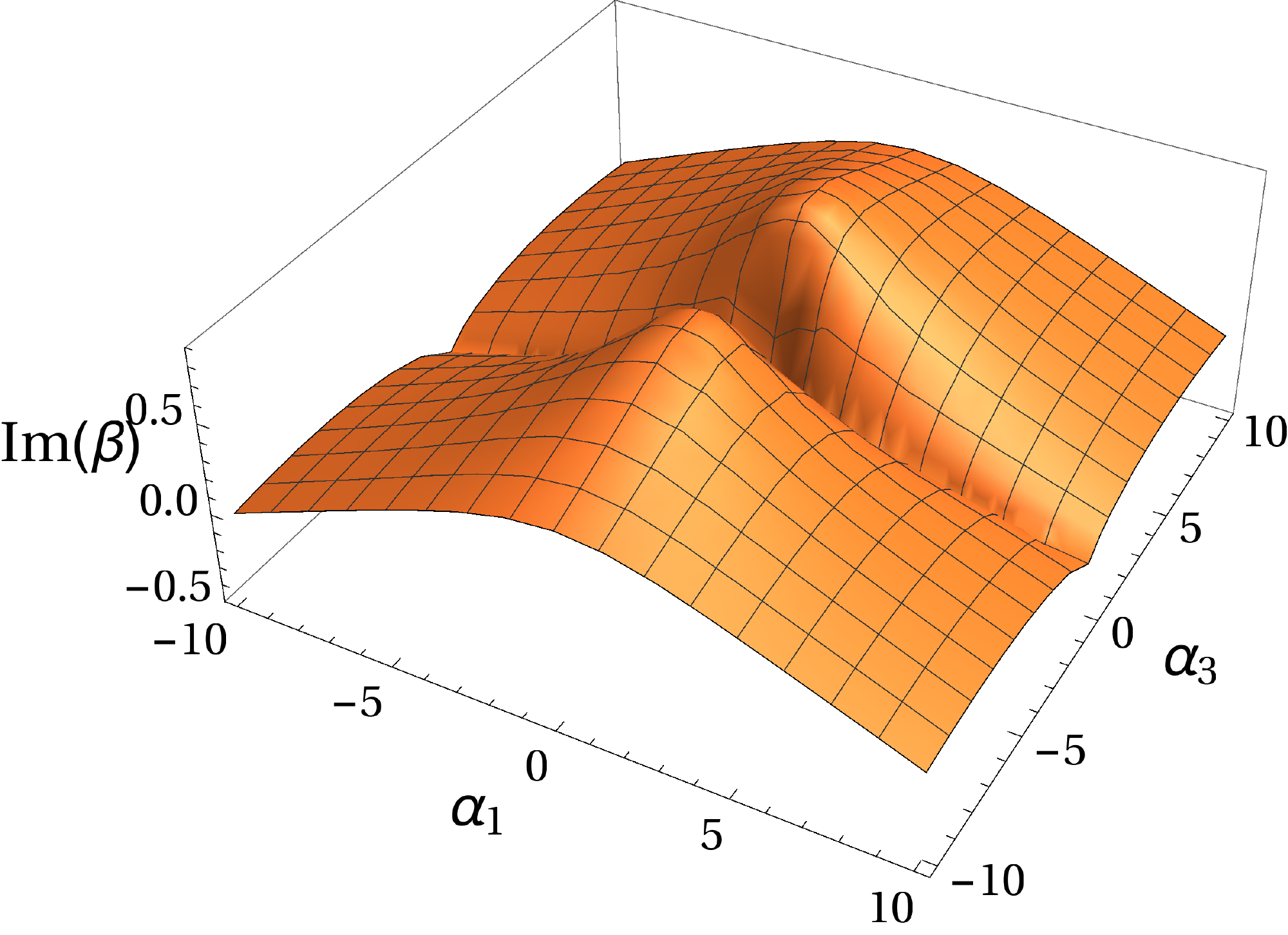}
  \caption{Variation of $Im(\beta)$ in three dimensions as a function of $\alpha_1$ and $\alpha_3$ for $Re = 10^3$ and $m/u_0 = 10$ for 
the Keplerian flow. $Im(\beta)_{max} = 0.896$.}
\label{fig:kep_wkb_3d_plot_v_10_re_10_3}
\end{figure}

\begin{figure}
 \includegraphics[width=\columnwidth]{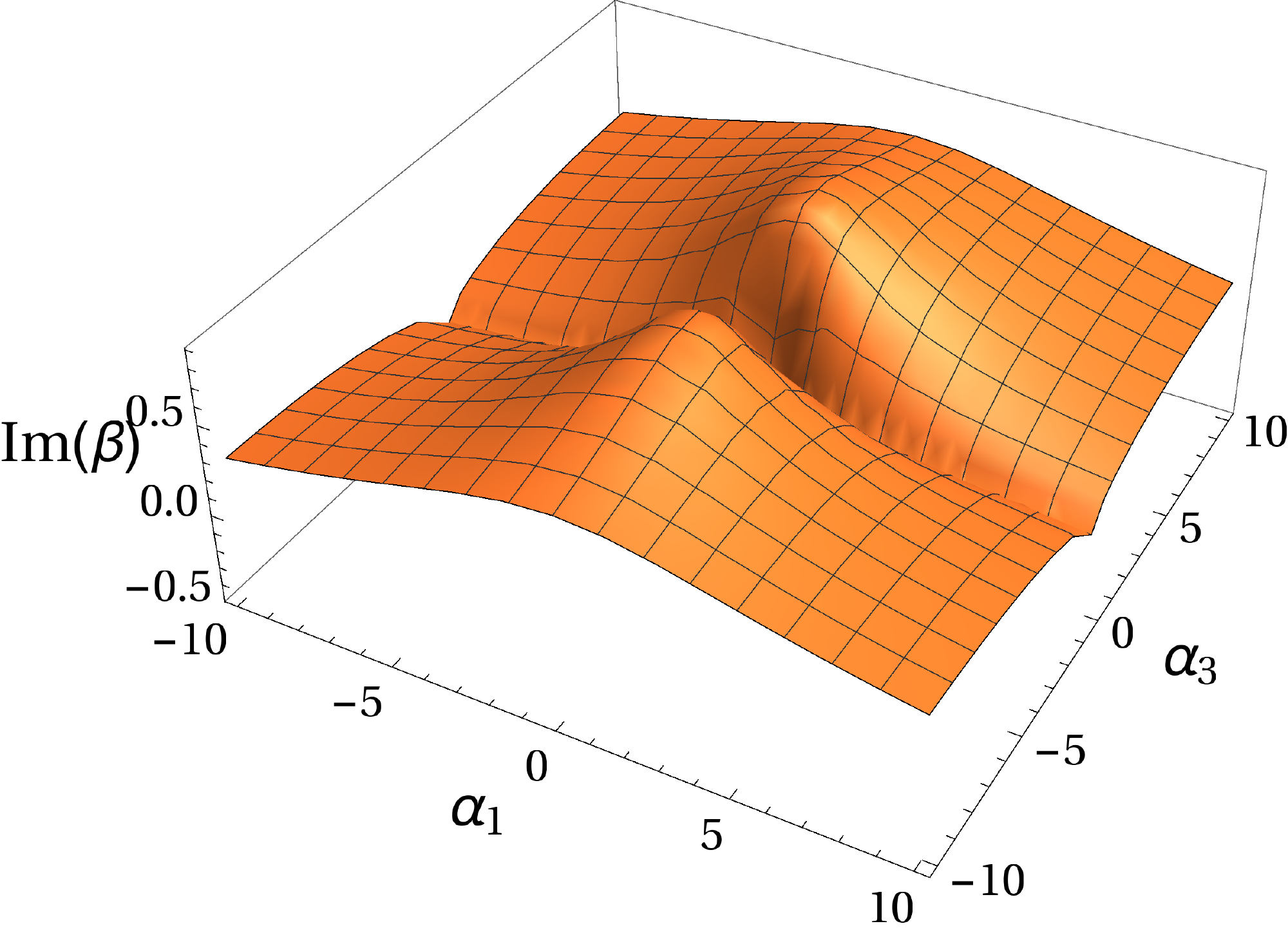}
  \caption{Variation of $Im(\beta)$ in three dimensions as a function of $\alpha_1$ and $\alpha_3$ for $Re = 10^4$ and $m/u_0 = 10$ for 
the Keplerian flow. $Im(\beta)_{max} = 0.91$.}
\label{fig:kep_wkb_3d_plot_v_10_re_10_4}
\end{figure}

Now if we fix $m/u_0$ and increase $Re$, 
it is expected that the value of $Im(\beta)$ increases. FIGs.~\ref{fig:kep_wkb_3d_plot_v_10_re_10}, \ref{fig:kep_wkb_3d_plot_v_10_re_10_2}, 
\ref{fig:kep_wkb_3d_plot_v_10_re_10_3} and \ref{fig:kep_wkb_3d_plot_v_10_re_10_4} depict the same. These four figures show the variation of 
$Im(\beta)$ as a function of $\alpha_1$ and $\alpha_3$ in three dimensions for $Re=10,\ 10^2,\ 10^3\ {\rm and}\ 10^4$ for $m/u_0 = 10$ in 
case of the Keplerian flow. $Im(\beta)_{max}$ is given in the caption corresponding to each figure to compare one with other. We make 
three dimensional plots for these cases to capture $Im(\beta)_{max}$, as it is not obvious from the contour plots, particularly from 
FIGs.~\ref{fig:keplarian_wkb_v_10_re_10_2} and \ref{fig:keplarian_wkb_v_10_re_10_10}. This fact becomes clear once we compare between 
FIGs.~\ref{fig:kep_wkb_v_10_re_10_4} and \ref{fig:kep_wkb_3d_plot_v_10_re_10_4}. From these four three dimensional figures and also from 
FIGs.~\ref{fig:keplarian_wkb_v_10_re_10_2}, \ref{fig:kep_wkb_v_10_re_10_4} and \ref{fig:keplarian_wkb_v_10_re_10_10}, it is clear that 
the increment of $Re$ for a fixed $m/u_0$ also increases the range of $\alpha_1$ and $\alpha_3$ which could
give rise to positive $Im(\beta)$ and hence instability in the system. To capture this particular fact, 
we zoom out the axes of the FIGs.~\ref{fig:keplarian_wkb_v_10_re_10_10} and \ref{fig:kep_wkb_v_10_re_10_4} as 
these two figures look almost similar if the bound on the axes is chosen from -10 to 10. Similarly,
FIGs.~\ref{fig:kep_wkb_3d_plot_v_10_re_10_3} and \ref{fig:kep_wkb_3d_plot_v_10_re_10_4} may apparently look
same, however they are not. If we check the fact that at which value of $Im(\beta)$, the surfaces of $Im(\beta)$ corresponding to these two 
figures cut the $Im(\beta)$ axis at $\alpha_1 = -10$, then we can be sure that they are not same. Apart from this, 
FIG.~\ref{fig:kep_wkb_3d_plot_v_10_re_10_3} shows that at $\alpha_1 = -10$, the surface of $Im(\beta)$ is downwards while the same for 
FIG.~\ref{fig:kep_wkb_3d_plot_v_10_re_10_4} is almost flat.

\begin{figure}
 \includegraphics[width=\columnwidth]{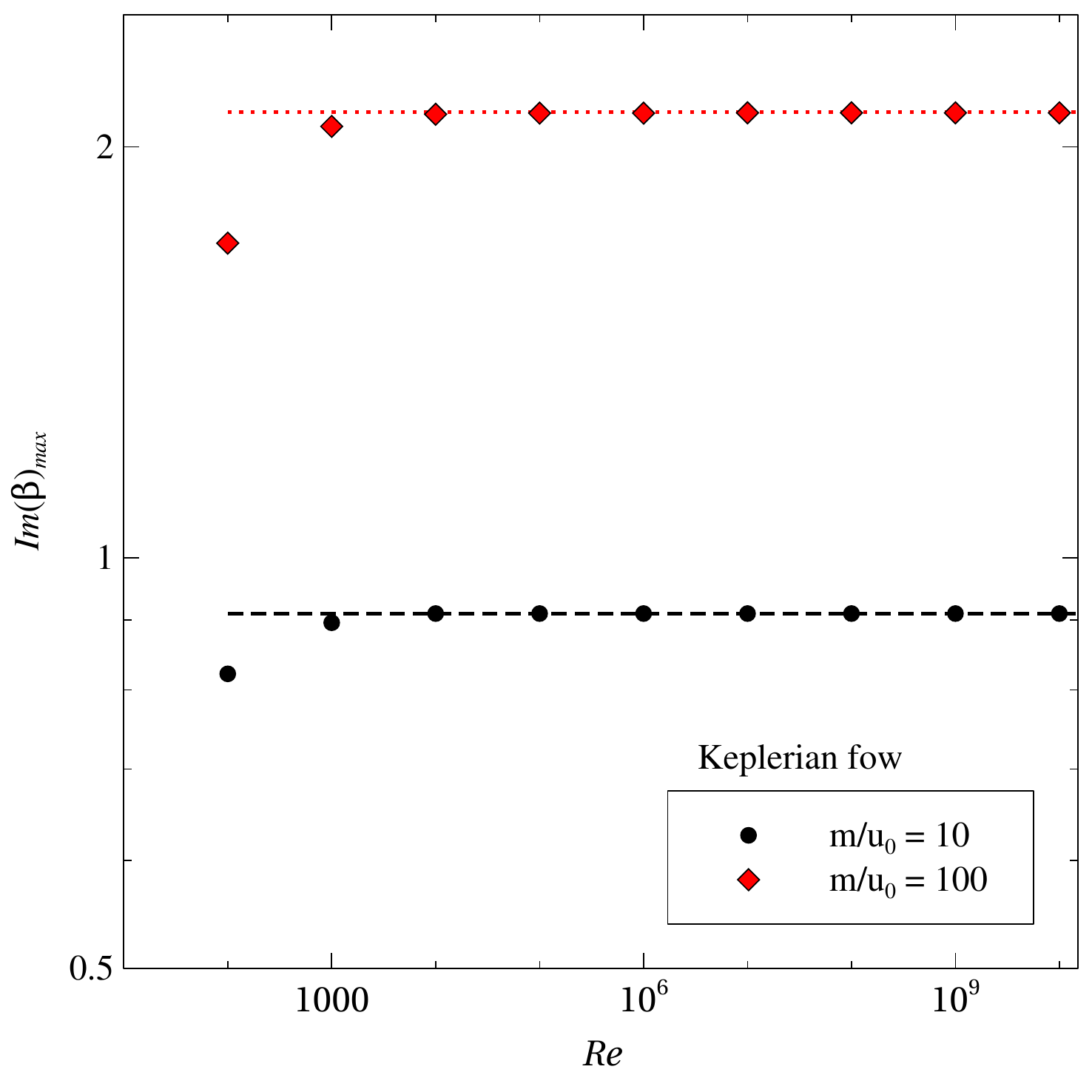}
  \caption{Variation of $Im(\beta)_{max}$ as a function of $Re$ for $m/u_0 = 10$ and $m/u_0 = 100$ for the Keplerian flow. The 
  dashed and dotted lines represent $Im(\beta)_{max} = 0.91$ and $2.12$ respectively.}
\label{fig:Im_beta_max_vs_Re}
\end{figure}

However, $Im(\beta)_{max}$ does not increase beyond 0.91, even if we increase 
$Re$ for $m/u_0 = 10$ for the Keplerian flow. It, therefore, looks like $Im(\beta)_{max}$ gets saturated at 0.91 at $Re=10^4$ and any further 
increment in $Re$ increases only the range of $\alpha_1$ and $\alpha_3$ that makes $Im(\beta)$ positive. This saturation of $Im(\beta)$ 
depends on $m/u_0$. FIG.~\ref{fig:kep_wkb_v_10_2_re_10_4} shows the variation of $Im(\beta)$ as a function of 
$\alpha_1$ and $\alpha_3$ for $Re = 10^4$ and $m/u_0 = 10^2$ in the case of Keplerian flow. In this case, $Im(\beta)_{max}$ is 2.12. 
Increment of $m/u_0$, therefore, increases the saturation in $Im(\beta)_{max}$. This situation is well-depicted in
FIG.~\ref{fig:Im_beta_max_vs_Re} which shows the variation of $Im(\beta)_{max}$ as a function of $Re$ for $m/u_0 = 10$ and $m/u_0 = 100$ 
for the Keplerian flow. In addition, the same figure also shows the saturation of $Im(\beta)_{max}$ for a fixed $m/u_0$. 
If we consider $\alpha_1 = 0$ in equation (\ref{eq:sol_beta}), we obtain the dispersion relations as shown by Nath \& 
Mukhopadhyay \cite{Nath_2016} in Figure 2. 

\subsection{Plane Couette Flow}
\label{subsec:plane_cou_flow}
For plane Couette flow, equation (\ref{eq:sol_beta}) becomes
\begin{eqnarray}
\begin{split}
\beta = -\frac{0.5 i}{3 \alpha _1^2+\alpha _3^2}\Bigl[\frac{m}{u_0}+\frac{14 \alpha _1^4}{Re}+\frac{12 \alpha _3^2 \alpha _1^2}{Re} & \\ 
+\frac{2 \alpha _3^4}{Re} -\frac{1}{Re}\Bigl(\frac{m^2 Re^2}{u_0^2} -\frac{8 \alpha _1^4 m Re}{u_0}+16\alpha _1^8\Bigr)^{\frac{1}{2}}\Bigr].
\end{split}
\label{eq:sol_beta_plane_cou}
\end{eqnarray}
It is quite obvious that $\beta$ is an imaginary quantity for plane Couette flow. To have instability, therefore, the quantity within the 
square bracket must be negative and this leads to the condition
\begin{equation}
\begin{split}
\frac{m}{u_0}<-\frac{45}{Re \left(9\text{$\alpha_1 $}^4+6\text{$\alpha_1 
$}^2 \text{$\alpha_3 $}^2+\text{$\alpha_3 $}^4\right)}\Bigl(\text{$\alpha_1 $}^8+1.867 \text{$\alpha_1 $}^6 \text{$\alpha_3 $}^2\\+1.111 
\text{$\alpha_1 $}^4 \text{$\alpha_3 $}^4+0.267 \text{$\alpha_1 $}^2 \text{$\alpha_3 $}^6+0.022 \text{$\alpha_3 $}^8\Bigr).
\end{split}
\label{eq:v_condition}
\end{equation}
$m/u_0$, therefore, has to be negative to have instability in plane Couette flow. If we make $\alpha_1 = 0$, the condition in equation 
(\ref{eq:v_condition}) becomes
\begin{equation}
 \frac{m}{u_0}<-\frac{\alpha_3^4}{Re},
\end{equation}
which was obtained by Nath \& Mukhopadhyay (2016) \cite{Nath_2016} for vertical perturbation.

\begin{figure}
 \includegraphics[width=\columnwidth]{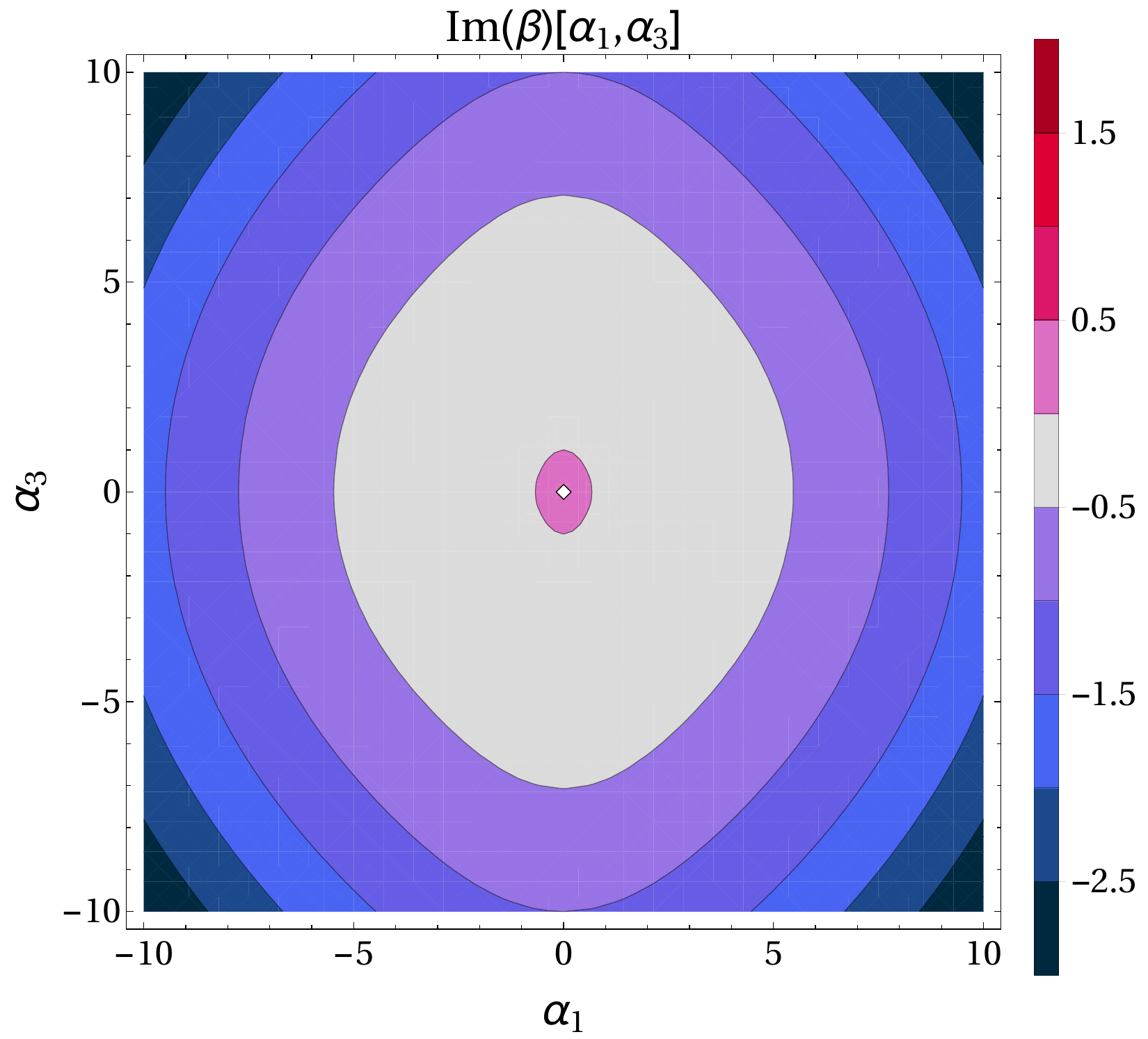}
  \caption{Variation of $Im(\beta)$ as a function of $\alpha_1$ and $\alpha_3$ for $Re = 10^{2}$ and $m/u_0 = -10^{-2}$ 
for plane Couette flow. At $\alpha_1 = \alpha_3 = 0$, $Im(\beta)\rightarrow \infty$. It is indicated by 
white point at the center of the plot.}
\label{fig:plane_cou_wkb_v_-10_-2_re_10_2}
\end{figure}

\begin{figure}
 \includegraphics[width=\columnwidth]{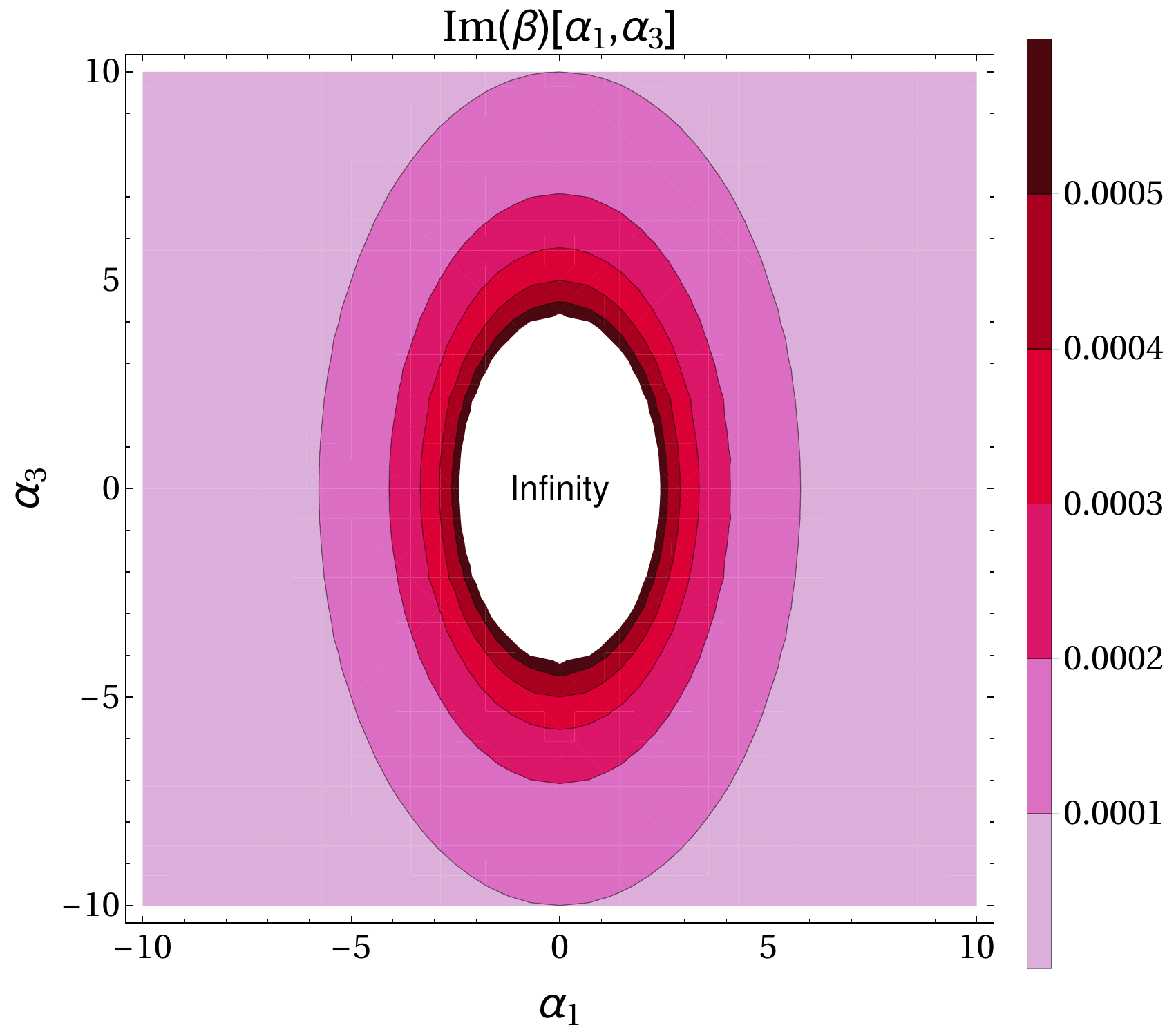}
  \caption{Variation of $Im(\beta)$ as a function of $\alpha_1$ and $\alpha_3$ for $Re = 10^{10}$ and $m/u_0 = 
-10^{-2}$ 
for plane Couette flow.}
\label{fig:plane_cou_wkb_v_-10_-2_re_10_10}
\end{figure}

\begin{figure}
 \includegraphics[width=\columnwidth]{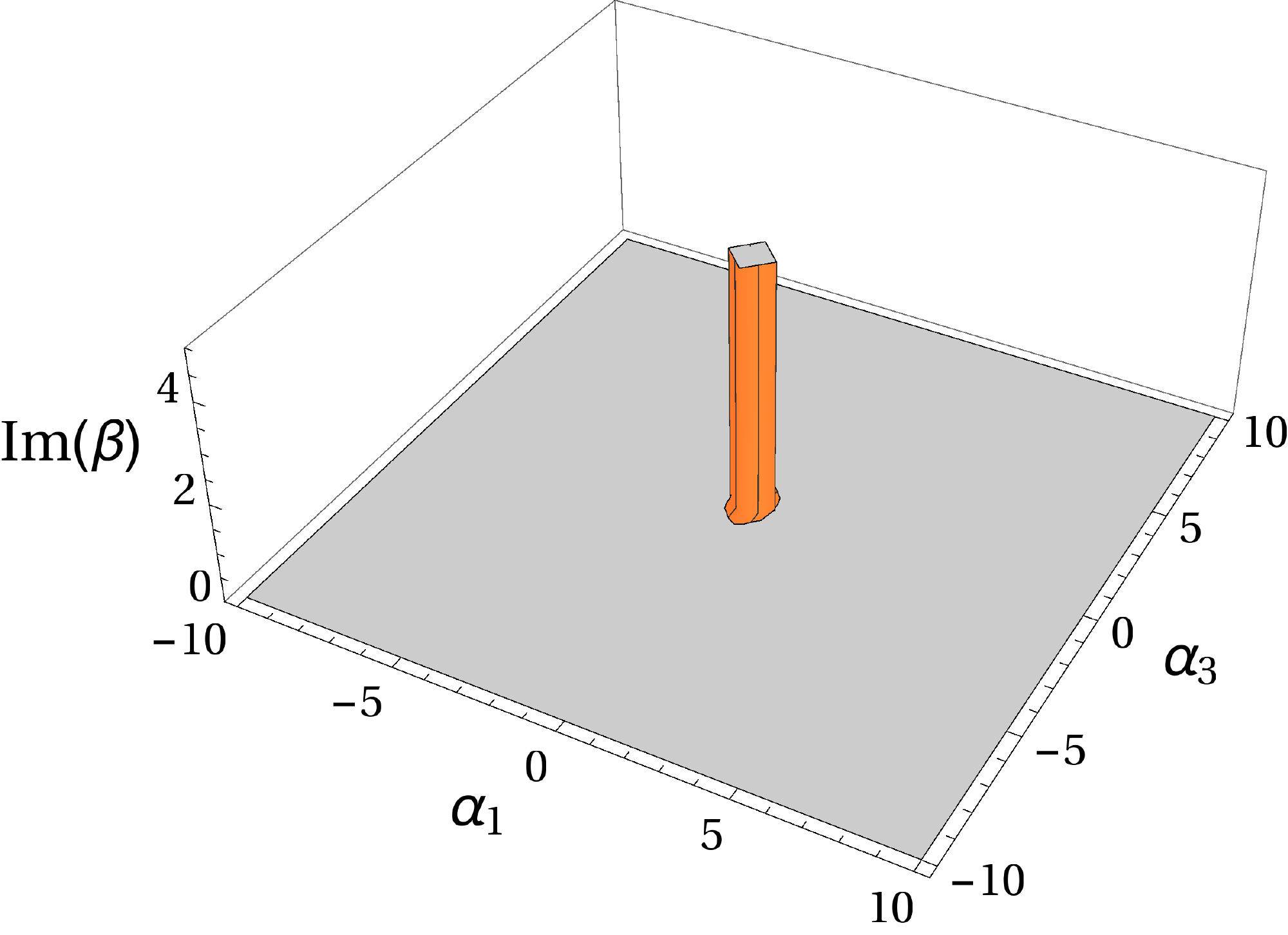}
  \caption{Variation of $Im(\beta)$ in three dimensions as a function of $\alpha_1$ and $\alpha_3$ for $Re = 10^{2}$ and $m/u_0 = -10^{-2}$ 
for plane Couette flow.}
\label{fig:plane_cou_wkb_3d_plot_v_-10_-2_re_10_2}
\end{figure}

\begin{figure}
 \includegraphics[width=\columnwidth]{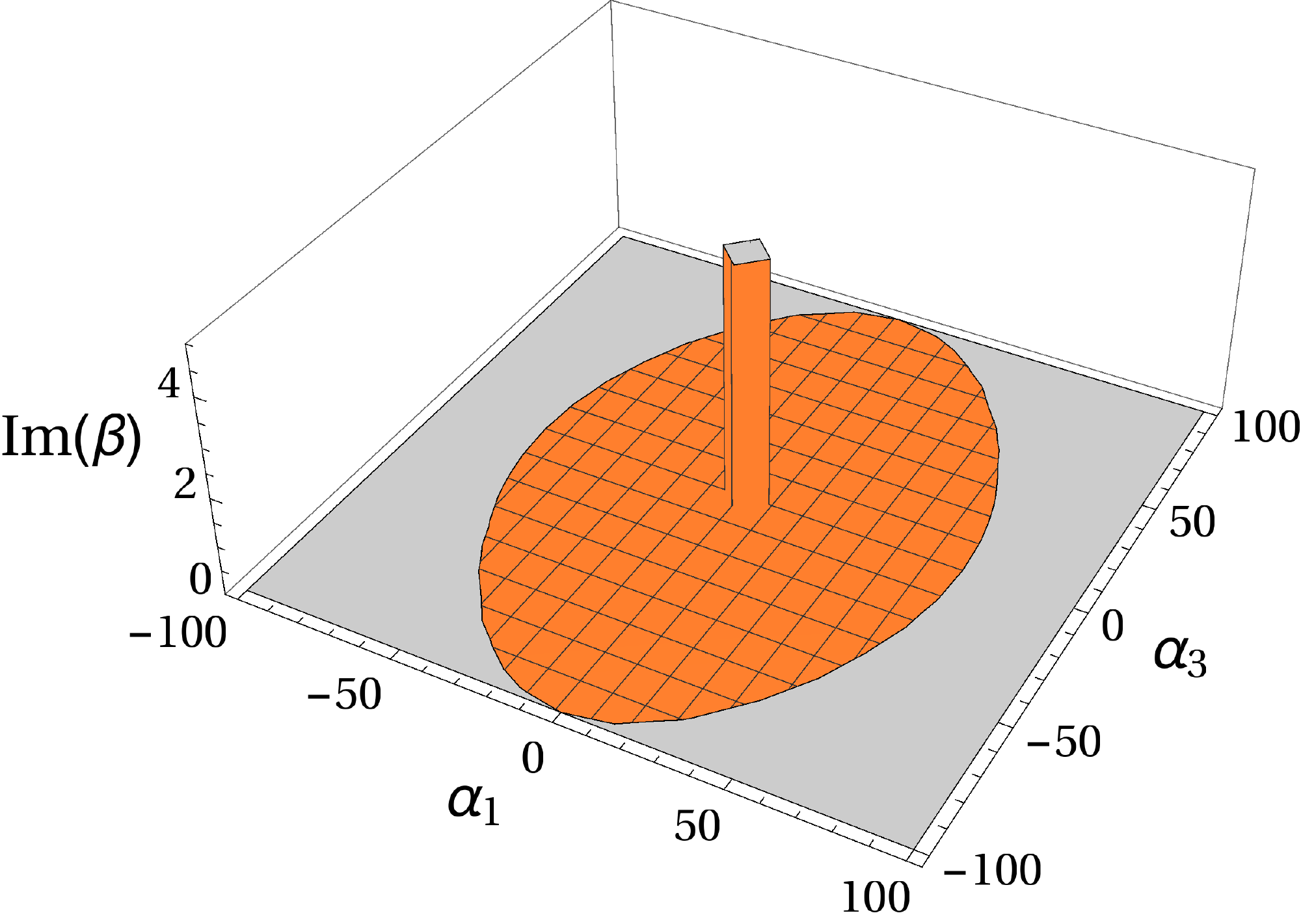}
  \caption{Variation of $Im(\beta)$ in three dimensions as a function of $\alpha_1$ and $\alpha_3$ for $Re = 10^{10}$ and $m/u_0 = 
-10^{-2}$ 
for plane Couette flow.}
\label{fig:plane_cou_wkb_3d_plot_v_-10_-2_re_10_10_alpha_100}
\end{figure}

\begin{figure}
 \includegraphics[width=\columnwidth]{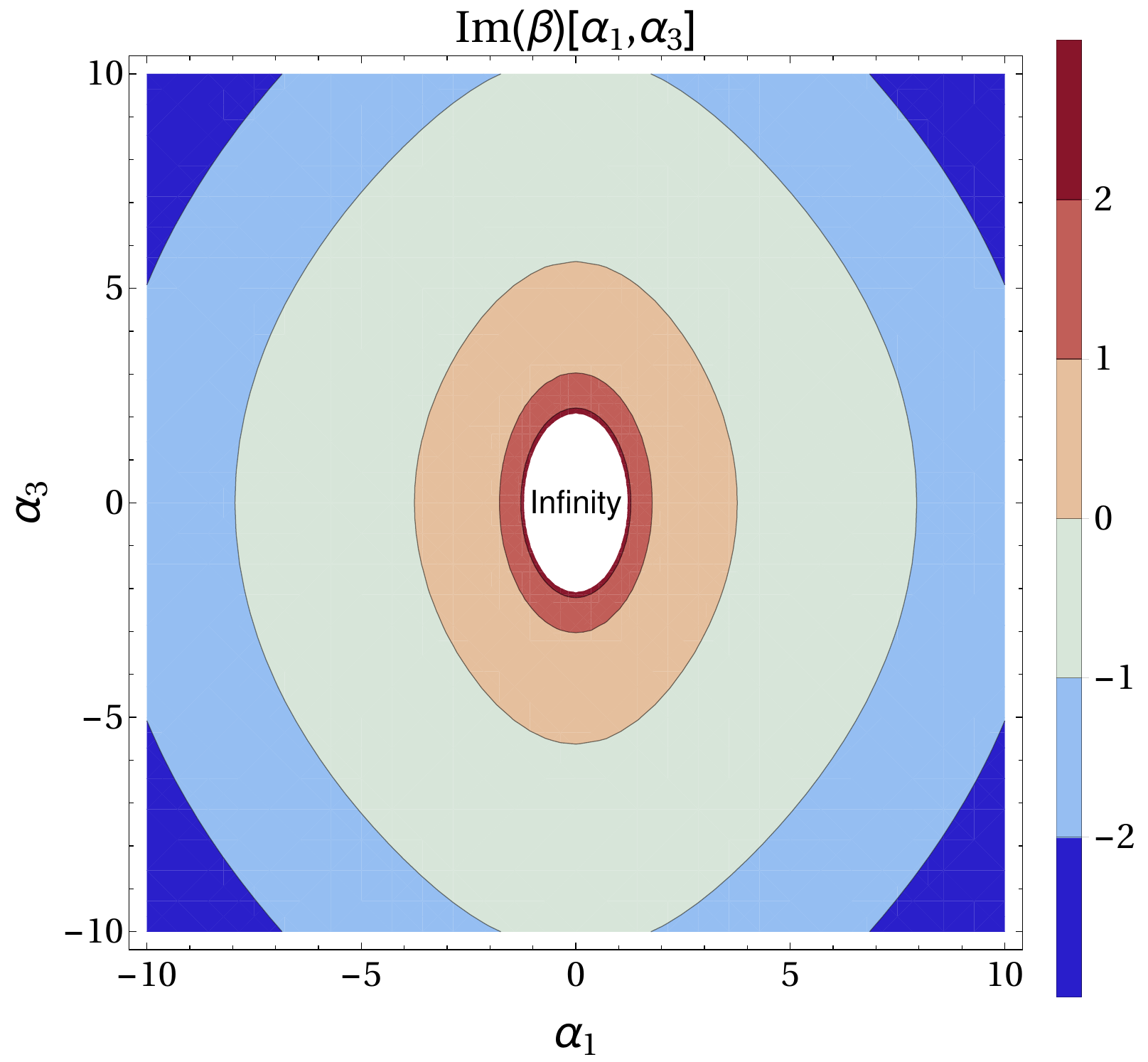}
  \caption{Variation of $Im(\beta)$ as a function of $\alpha_1$ and $\alpha_3$ for $Re = 10^{2}$ and $m/u_0 = -10$ 
for plane Couette flow.}
\label{fig:plane_cou_wkb_v_-10_re_10_2}
\end{figure}

\begin{figure}
 \includegraphics[width=\columnwidth]{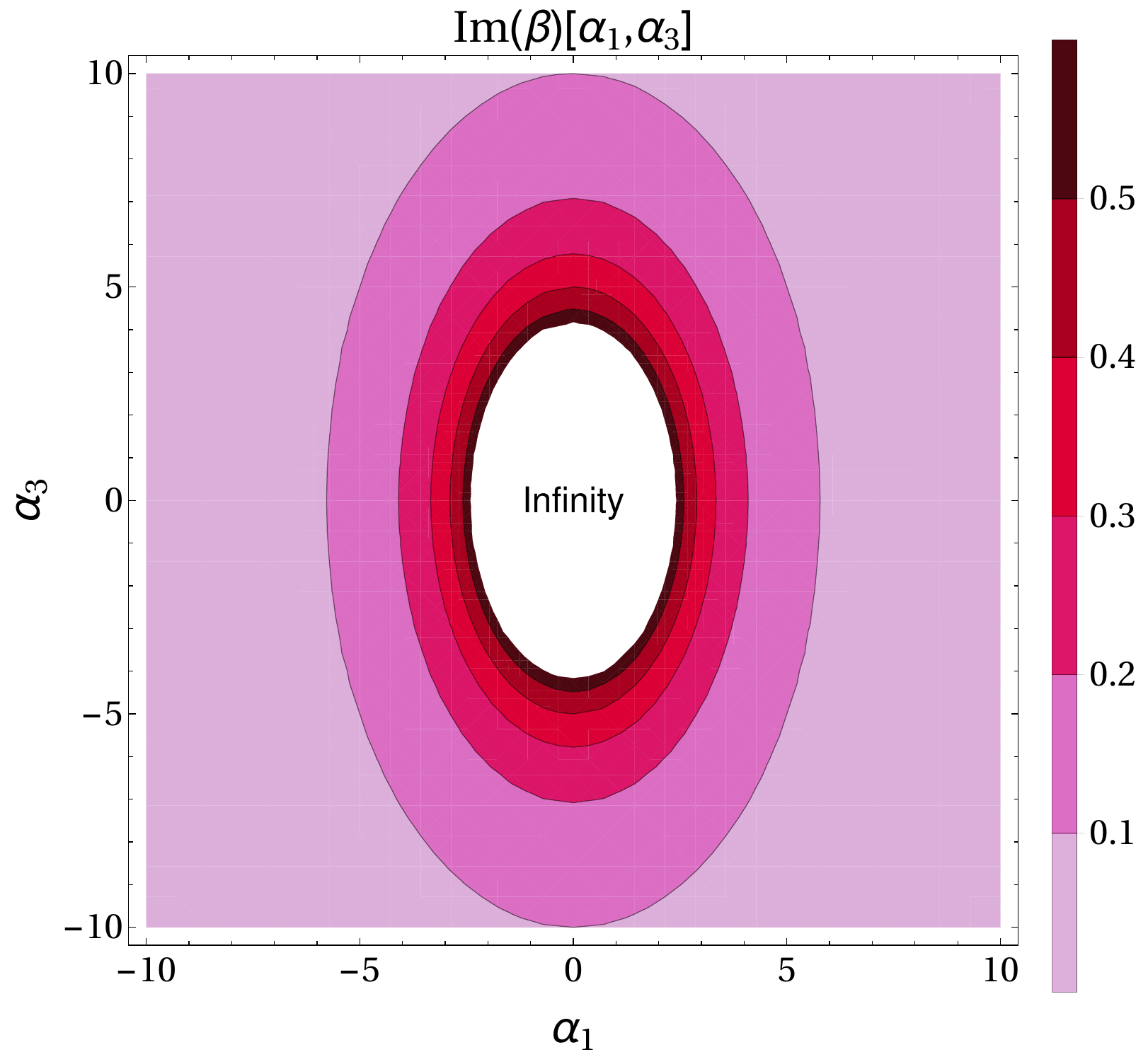}
  \caption{Variation of $Im(\beta)$ as a function of $\alpha_1$ and $\alpha_3$ for $Re = 10^{10}$ and $m/u_0 = -10$ 
for plane Couette flow.}
\label{fig:plane_cou_wkb_v_-10_re_10_10}
\end{figure}

From equation (\ref{eq:sol_beta_plane_cou}), it is obvious that $Im(\beta)$ blows up at $\alpha_1 = \alpha_3 = 0$. The 
color bars in the 
contour plots corresponding to plane Couette flow, therefore, have different meaning than indicating the value of $Im(\beta)$. They, 
rather, indicate the range of $\alpha_1$ and $\alpha_3$ within which $Im(\beta)$ has positive value, i.e. flow is 
unstable. 

We use the same color codes for the contour plots in
FIGs.~\ref{fig:plane_cou_wkb_v_-10_-2_re_10_2}, \ref{fig:plane_cou_wkb_v_-10_-2_re_10_10}, 
\ref{fig:plane_cou_wkb_v_-10_re_10_2} and \ref{fig:plane_cou_wkb_v_-10_re_10_10}, as used in \S 
\ref{subsec:kep_flow}. However, we use grayish color to indicate the transition from the positive to 
negative of $Im(\beta)$. As $\alpha_1, \alpha_3 \rightarrow 0$, $Im(\beta)\rightarrow \infty$. The region where 
$\alpha_1, \alpha_3 \rightarrow 0$, therefore, cannot be captured in the contour plots. This region, therefore, is 
covered with white color by default. However, to avoid any confusion, we 
mention `Infinity' inside this region 
wherever possible, otherwise we mention it in the corresponding captions.

FIGs.~\ref{fig:plane_cou_wkb_v_-10_-2_re_10_2} and \ref{fig:plane_cou_wkb_v_-10_-2_re_10_10} show the 
variation of $Im(\beta)$ as a 
function of $\alpha_1$ and $\alpha_3$ for $m/u_0 = -10^{-2}$, $Re = 10^2$ and $10^{10}$ respectively for plane Couette flow. There is 
no negative $Im(\beta)$ in FIG.~\ref{fig:plane_cou_wkb_v_-10_-2_re_10_10} within the ranges of $\alpha_1$ and $\alpha_3$. On the 
contrary, there are negative values of $Im(\beta)$ in FIG.~\ref{fig:plane_cou_wkb_v_-10_-2_re_10_2} within the same range of $\alpha_1$ and 
$\alpha_3$. For the same $m/u_0$ (which is also very small here), therefore, the increment in $Re$ increases the range of $\alpha_1$ and 
$\alpha_3$ which gives rise to positive $Im(\beta)$ and hence increases the chance of making the system unstable. 
FIGs.~\ref{fig:plane_cou_wkb_3d_plot_v_-10_-2_re_10_2} and \ref{fig:plane_cou_wkb_3d_plot_v_-10_-2_re_10_10_alpha_100} make this point 
even clearer. These two figures represent the variation of $Im(\beta)$ ($\geq 0$) in three dimensions as a function of $\alpha_1$ and 
$\alpha_3$ for $Re = 10^{2}$ and $10^{10}$ respectively for $m/u_0 = -10^{-2}$ for plane Couette flow.

It is also expected that if we increase the magnitude of $m/u_0$, the system becomes more unstable as in the case of Keplerian flow. This 
phenomenon also happens here but in different way. FIGs.~\ref{fig:plane_cou_wkb_v_-10_re_10_2} and \ref{fig:plane_cou_wkb_v_-10_re_10_10} 
show the variation of $Im(\beta)$ as a function of $\alpha_1$ and $\alpha_3$ for $Re = 10^{2}$ and $10^{10}$ respectively and 
for $m/u_0 = -10$ for plane Couette flow. However, if we compare carefully FIG.~\ref{fig:plane_cou_wkb_v_-10_-2_re_10_2}
(or FIG.~\ref{fig:plane_cou_wkb_v_-10_-2_re_10_10}) with FIG.~\ref{fig:plane_cou_wkb_v_-10_re_10_2}
(or FIG.~\ref{fig:plane_cou_wkb_v_-10_re_10_10}), we see that FIG.~\ref{fig:plane_cou_wkb_v_-10_re_10_2} 
(or FIG.~\ref{fig:plane_cou_wkb_v_-10_re_10_10}) has a larger range of $\alpha_1$ and 
$\alpha_3$ to give rise to positive $Im(\beta)$.

\section{Argand diagram}
\label{sec:arg_dia}
\begin{figure}
 \includegraphics[width=\columnwidth]{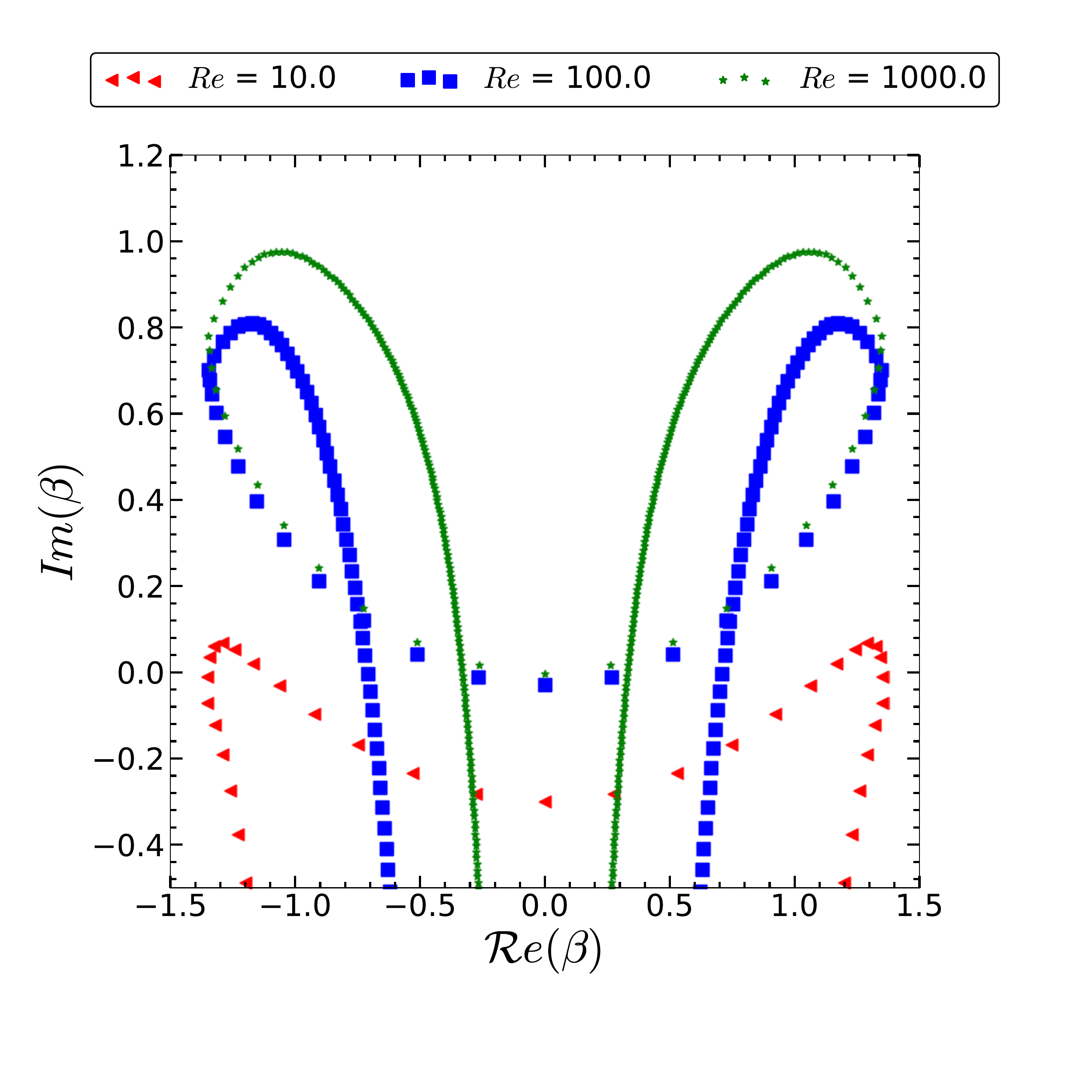}
  \caption{Argand diagram for $Re = 10$, 100 and 1000 for $m/u_0 = 10$ and $\alpha_1 = 1.0$ for the Keplerian flow. For each $Re$, 
$\alpha_{3}$ is varied from $-2000$ to 2000.}
\label{fig:WKB_kep_argand_re_10_1_2_3_v_10}
\end{figure}

\begin{figure}
 \includegraphics[width=\columnwidth]{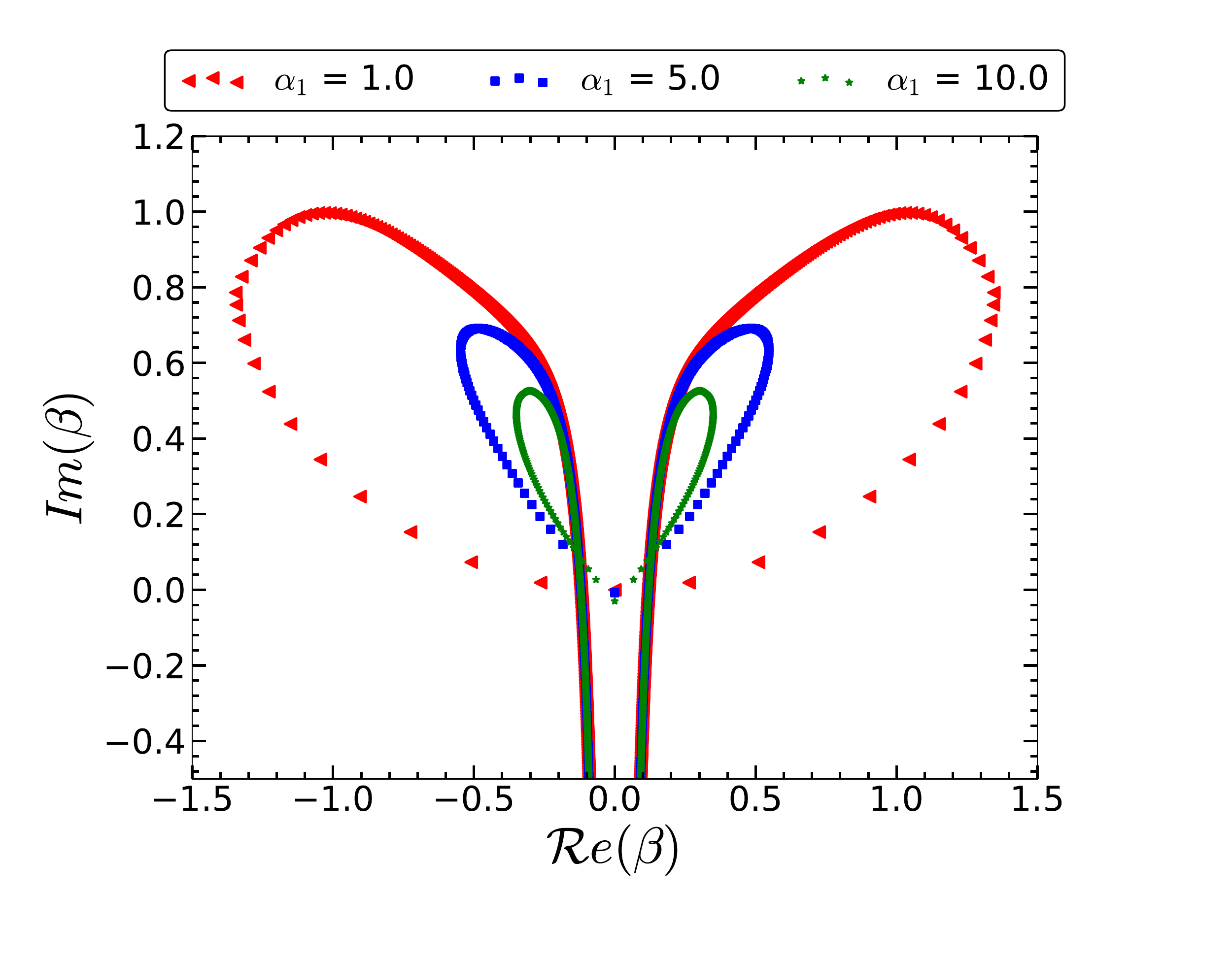}
  \caption{Argand diagram for $Re = 10^{4}$, $m/u_0 = 10$ and for $\alpha_1 = 1.0$, 5.0 and 10.0 for the Keplerian flow.}
\label{fig:WKB_kep_argand_re_10_4_v_10}
\end{figure}

The time variation of the perturbations is given by
\begin{equation}
 u, \zeta\sim e^{-i\mathcal{R}e(\beta) t} e^{Im(\beta) t}.
\end{equation} 
In \S\ref{sec:dis_rel}, we show that 
$Im(\beta)$ has positive value within a certain range of $\alpha_1$ and $\alpha_3$. For those values of $\alpha_1$ and $\alpha_3$, 
therefore, $e^{(Im(\beta)) t}$ increases exponentially with time. On the other hand, $e^{-i(\mathcal{R}e(\beta)) t}$ is oscillatory in 
time. The real part of the temporal variation of the perturbation is
\begin{equation}
 \mathcal{R}e(u), \mathcal{R}e(\zeta) \sim \cos(\mathcal{R}e(\beta) t) e^{Im(\beta) t}.
\end{equation}
Here, we observe the variation of $Im(\beta)$ as a function of $\mathcal{R}e(\beta)$. FIG.~\ref{fig:WKB_kep_argand_re_10_1_2_3_v_10} shows 
Argand diagrams for $Re = 10,\ 10^2,\ 10^3$ for fixed $\alpha_1$ ($=1$) and $m/u_0 = 10$ by varying $\alpha_3$ in case of the Keplerian 
flow. We observe that $Im(\beta)_{max}$, i.e. the maximum growth rate increases as we increase $Re$ for a 
fixed $m/u_0$.
FIG.~\ref{fig:WKB_kep_argand_re_10_4_v_10} shows the Argand diagrams for $Re = 10^{4}$, $m/u_0 = 10$ 
and for $\alpha_1$ = 1.0, 5.0 and 10.0 for the Keplerian flow, where for each $\alpha_1$, we vary $\alpha_3$ from $-2000$ to 2000. From 
FIG.~\ref{fig:WKB_kep_argand_re_10_4_v_10}, it is clear that as we decrease $\alpha_1$, $Im(\beta)_{max}$ increases. For smaller 
$\alpha_1$, therefore, the system becomes unstable at smaller time and plausibly becomes turbulent for those $\alpha_1$ first.

\begin{figure}
 \includegraphics[width=\columnwidth]{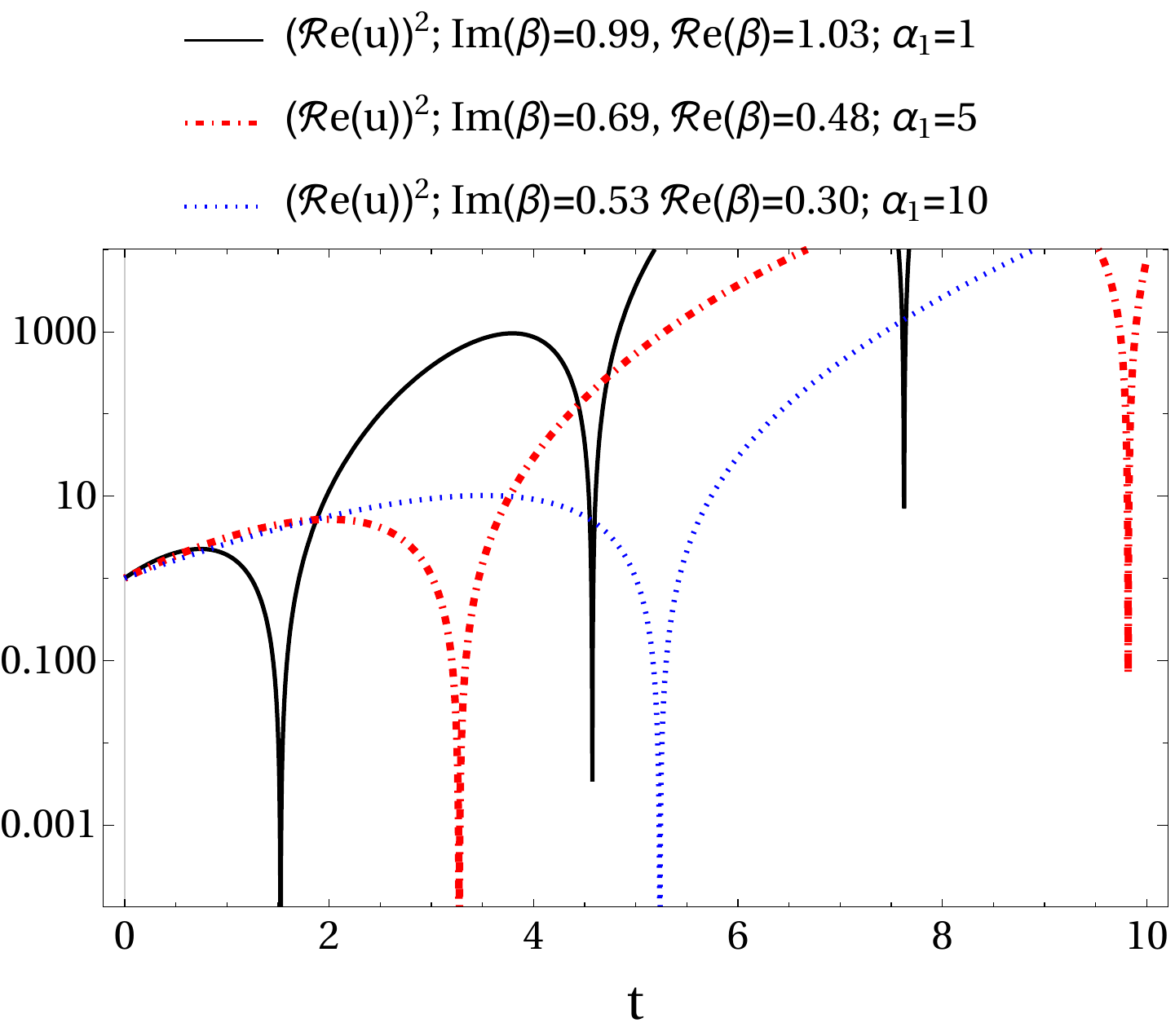}
  \caption{Variation of $(\mathcal{R}e(u))^2$ as a function of time, for $\mathcal{R}e(\beta)_{max}$ and 
$Im(\beta)_{max}$ from FIG.~\ref{fig:WKB_kep_argand_re_10_4_v_10} corresponding to $\alpha_1 = 1.0, 5.0\ {\rm and}\ 10.0$.}
\label{fig:f_t_squre}
\end{figure}

The phenomenon of increment in the maximum growth rate with decreasing $\alpha_1$ is described through 
the energy of perturbations in FIG.~\ref{fig:f_t_squre}. Here, $(\mathcal{R}e(u))^2$ represents the temporal 
evolution of energy corresponding to 
the $x$-component of the perturbed velocity field for $Im(\beta)_{max}$ and $\mathcal{R}e(\beta)_{max}$ 
corresponding to three different $\alpha_1$, $Re=10^4$ for $m/u_0 = 10$ in 
the case of the Keplerian flow. 
In FIG.~\ref{fig:f_t_squre}, the 
maximum value along vertical axis is $10^4$. We consider this value to be the limit of linearity following 
\cite{man_2005}. We notice that the higher $Im(\beta)_{max}$ has higher 
$\mathcal{R}e(\beta)_{max}$, i.e., the higher growth rates have the higher frequency.

\begin{figure}
 \includegraphics[width=\columnwidth]{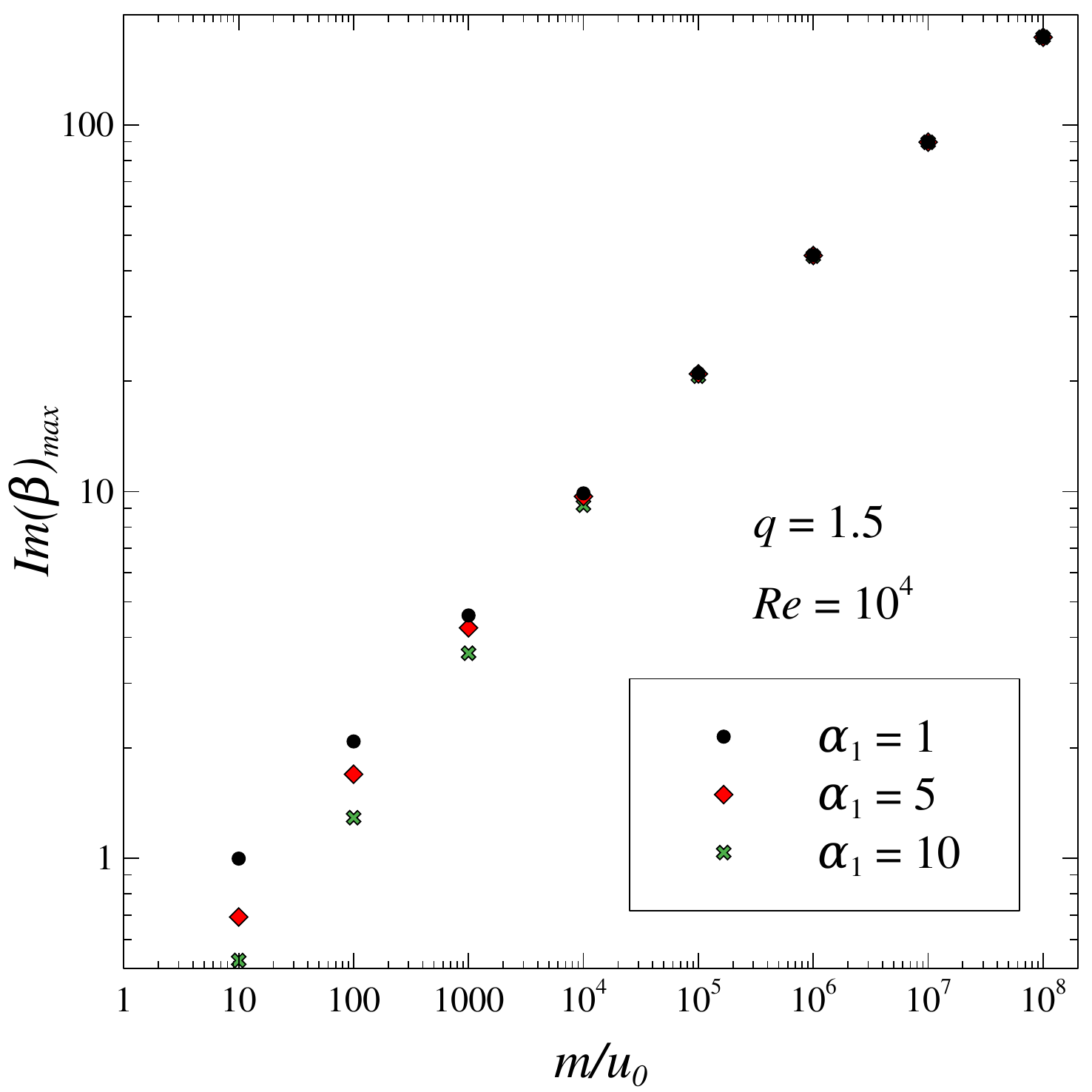}
  \caption{Variation of $Im(\beta)_{max}$ as a function of $m/u_0$, for $\alpha_1 = 1, 5, 10$ and $Re = 
10^4$ for the Keplerian flow.}
\label{fig:v_vs_growth_rates}
\end{figure}

It is always interesting to check what happens to the $Im(\beta)_{max}$  if $m/u_0$ 
increases for the same $Re$. FIG.~\ref{fig:v_vs_growth_rates} shows the variation of $Im(\beta)_{max}$ as a function of 
$m/u_0$, for $\alpha_1 = 1,\ 5,\ 10$ for $Re = 10^4$ for the Keplerian flow. Here we notice that $Im(\beta)_{max}$ increases 
as we increase $m/u_0$. However, at larger $m/u_0$, the $Im(\beta)_{max}$ becomes almost independent on $\alpha_1$. At 
higher
$m/u_0$, the extra force and $Re$ almost completely take control of the system of a fixed $q$. This phenomenon can be 
explained from the equation (\ref{eq:sol_beta}). At large $m/u_0$, the equation (\ref{eq:sol_beta}) becomes 
\begin{eqnarray}
 \begin{split}
\beta \sim -\frac{0.5 i}{3 \alpha _1^2 q+\alpha _3^2 q}\Bigl[\frac{m q}{u_0} -\frac{1}{Re}\Bigl(\frac{24i \alpha _3 
\alpha _1^2 m q Re^2}{u_0}\\ +\frac{8i \alpha _3^3 
m q Re^2}{u_0} +\frac{m^2 q^2 Re^2}{u_0^2} -\frac{8 \alpha _1^4 m q^2 Re}{u_0}\Bigr)^{\frac{1}{2}}\Bigr].
\end{split}
\label{eq:sol_beta_kep_large_m}
\end{eqnarray}
We obtain equation (\ref{eq:sol_beta_kep_large_m}) from equation (\ref{eq:sol_beta}) by retaining the terms that involve with
$m/u_0$ as the magnitude of other terms become negligible compared to those involving with $m/u_0$. From equation 
(\ref{eq:sol_beta_kep_large_m}), it is evident that as $m/u_0$ increases, the effect of $\alpha_1$ on $\beta$ and, hence, 
$Im(\beta)$ decreases.

\begin{figure}
\includegraphics[width=\columnwidth]{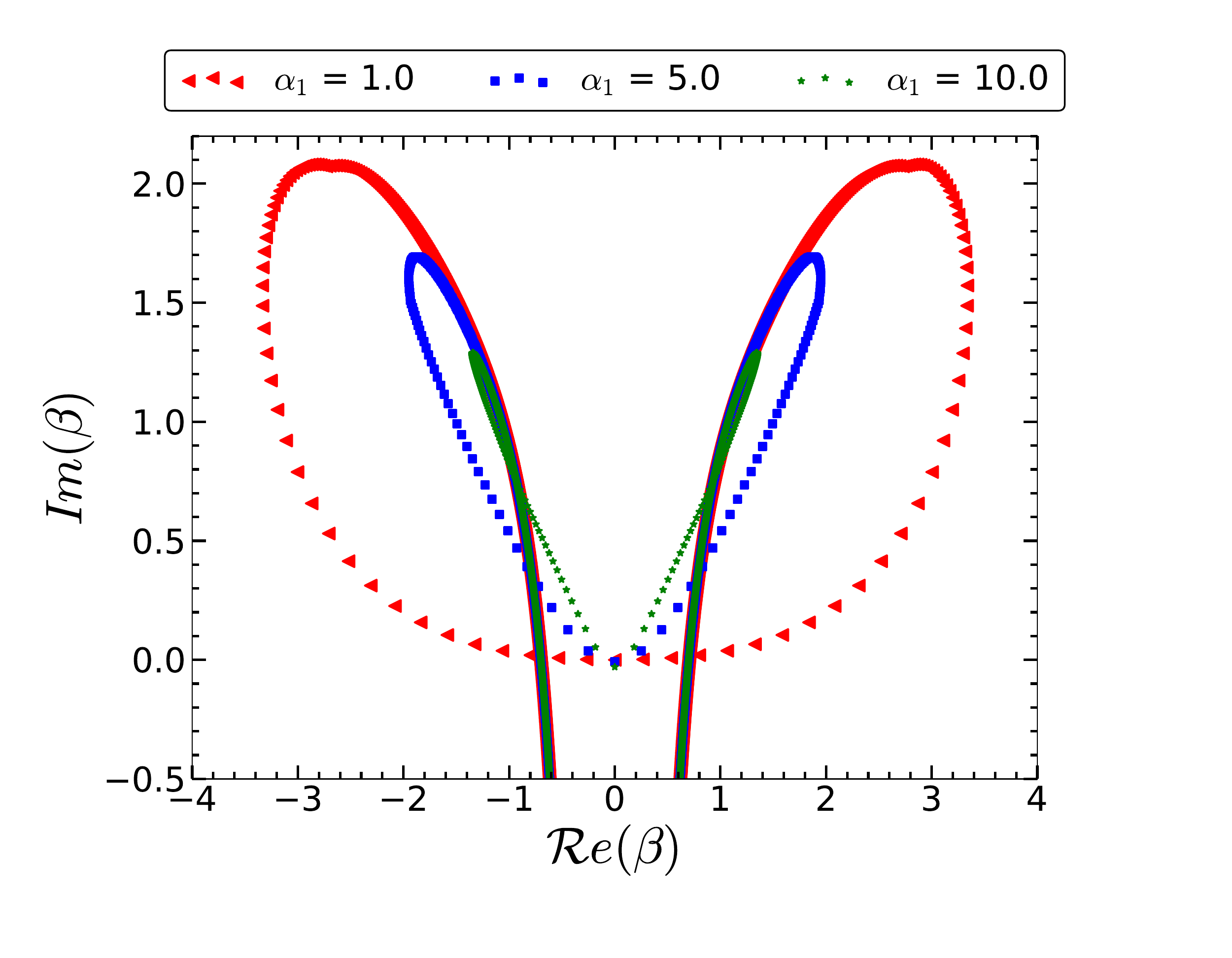}
  \caption{Argand diagram for $Re = 10^{4}$, $m/u_0 = 10^2$ and for $\alpha_1 = 1.0$, 5.0 and 10.0 for the Keplerian flow.}
\label{fig:WKB_kep_argand_re_10_4_v_10_2}
\end{figure}

\begin{figure}
\includegraphics[width=\columnwidth]{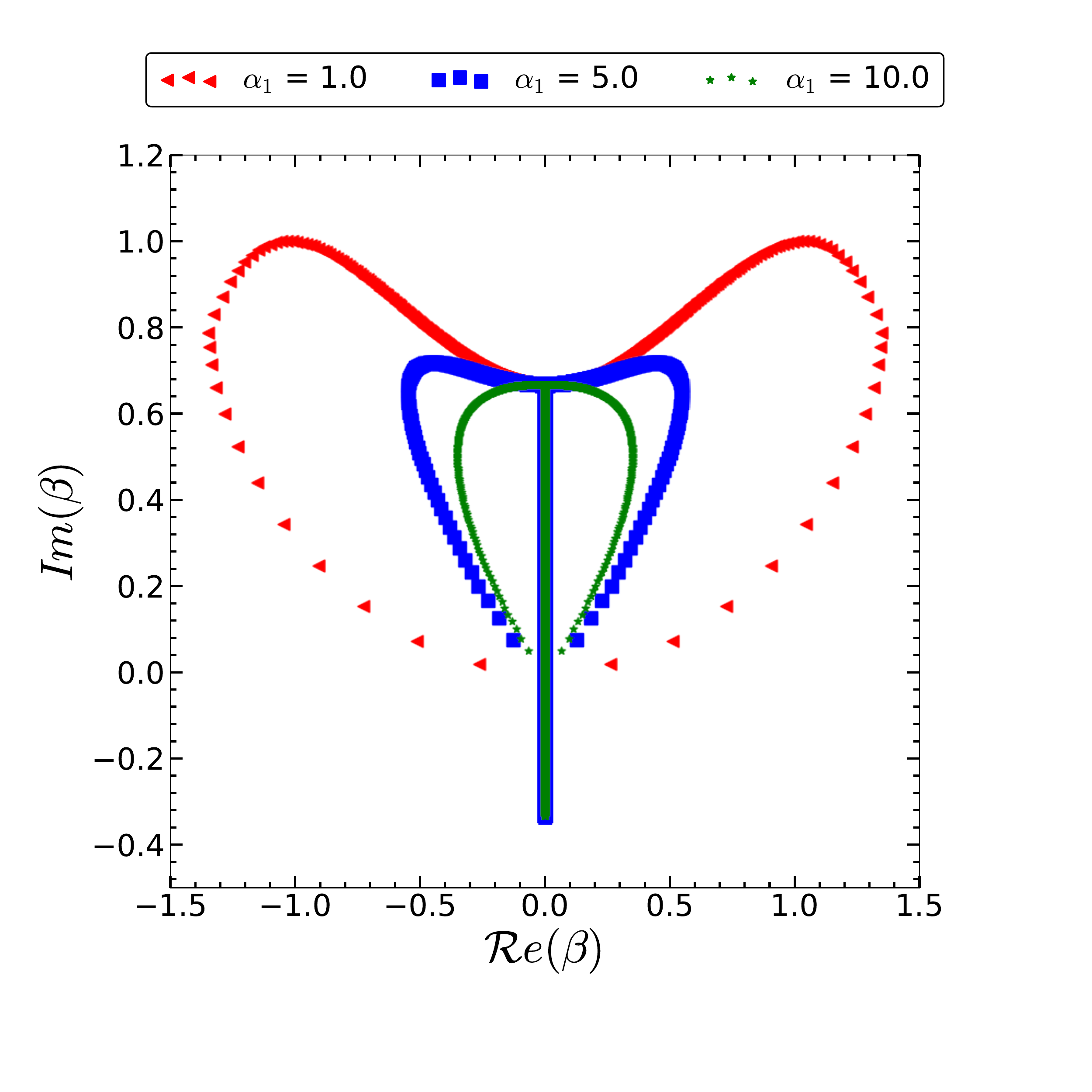}
  \caption{Argand diagram for $Re = 10^{10}$, $m/u_0 = 10$ and for $\alpha_1 = 1.0$, 5.0 and 10.0 for the Keplerian flow.}
\label{fig:WKB_kep_argand_re_10_10_v_10}
\end{figure}

\begin{figure}
\includegraphics[width=\columnwidth]{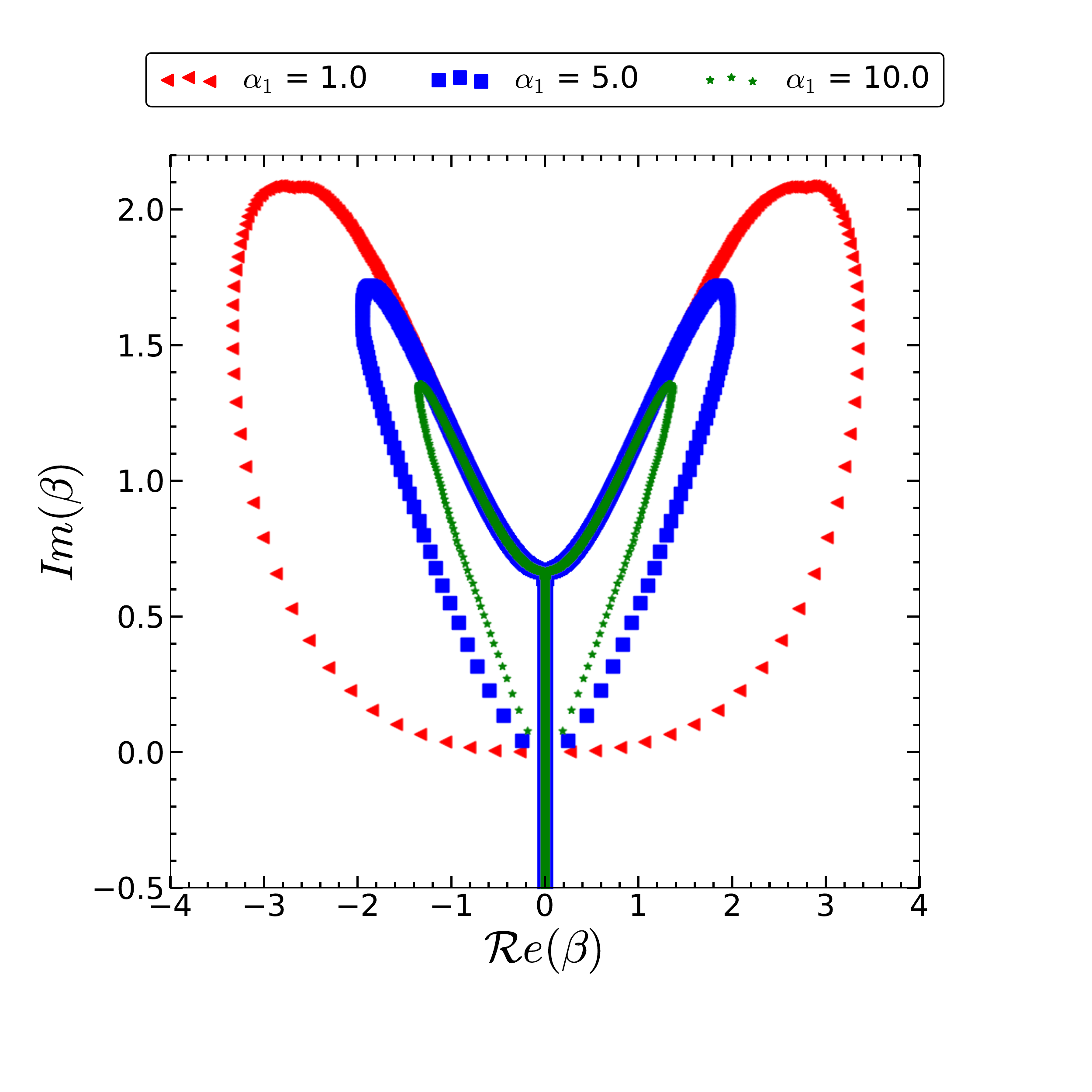}
  \caption{Argand diagram for $Re = 10^{10}$, $m/u_0 = 10^2$ and for $\alpha_1 = 1.0$, 5.0 and 10.0 for the Keplerian flow.}
\label{fig:WKB_kep_argand_re_10_10_v_10_2}
\end{figure}

To make the study complete, we should have enough comparison among Argand diagrams like 
FIG.~\ref{fig:WKB_kep_argand_re_10_4_v_10} but with different $m/u_0$ and $Re$. FIG.~\ref{fig:WKB_kep_argand_re_10_4_v_10_2} represents the Argand diagrams for $Re = 10^4$ and $m/u_0 = 10^2$ for three values of 
$\alpha_1$ mentioned in the figure where $\alpha_3$ is varied from $-2000$ to 2000 for each value of $\alpha_1$. Here we see that, 
$Im(\beta)_{max}$ and $\mathcal{R}e(\beta)_{max}$ for three different $\alpha_1$ are greater than those for $m/u_0 = 10$. We, therefore, 
confirm that as $m/u_0$ increases the value of $Im(\beta)_{max}$ and $\mathcal{R}e(\beta)_{max}$ also increase. 
FIGs.~\ref{fig:WKB_kep_argand_re_10_10_v_10} and \ref{fig:WKB_kep_argand_re_10_10_v_10_2} show the Argand diagrams for $Re = 10^{10}$ but 
for $m/u_0 = 10$ and $10^2$ respectively for three different $\alpha_1$ as shown in the corresponding figures and for each $\alpha_1$, we 
vary $\alpha_3$ from $-100000$ to 100000. If we compare between FIGs.~\ref{fig:WKB_kep_argand_re_10_4_v_10} and 
\ref{fig:WKB_kep_argand_re_10_10_v_10} (and also between FIGs.~\ref{fig:WKB_kep_argand_re_10_4_v_10_2} and 
\ref{fig:WKB_kep_argand_re_10_10_v_10_2}), we notice that $Im(\beta)_{max}$ and $\mathcal{R}e(\beta)_{max}$ do not change as we increase 
$Re$ for a fixed $m/u_0$ ($=10$), but the range of $\alpha_1$, that gives rise to positive $Im(\beta)$, does increase, as the positive area 
under the curve increases with increasing $Re$.

\section{Comparison of various timescales}
\label{sec:discusion}
In \S\ref{sec:dis_rel}, we obtain the dispersion relation for the linear perturbation in the presence of Coriolis force and extra 
force for the Keplerian flow as well as plane Couette flow. It shows that there is a range of wave vectors in which $Im(\beta)$ is 
positive. On the other hand, we also see the presence of temporal oscillation in the linear perturbation due to the presence of 
$\mathcal{R}e(\beta)$ in \S\ref{sec:arg_dia}. It, therefore, is important to compare the time period of the temporal growth of the 
perturbation with the infall time scale. To calculate the infall time scale of the fluid parcel, we need the radial component of velocity of 
the flow in the Keplerian disk and it is given by (see e.g. \cite{Frank_2002})
\begin{eqnarray}
\begin{split}
 v_r(R) = 2.7\times10^4\times 
\alpha_s^{\frac{4}{5}}\Bigl(\frac{\dot{M}}{10^{16}}\Bigr)^{\frac{3}{10}}\Bigl(\frac{M}{M\textsubscript{\(\odot\)}}\Bigr)^{-\frac{1}{4}}\\
\Bigl(\frac{R}{10^{10}}\Bigr)^{-\frac{1}{4}}\Bigl(1-\left(\frac{R_*}{R}\right)^{1/2}\Bigr)^{-7/10} {\rm cm\ s}^{-1},
\end{split}
\end{eqnarray}
where $\alpha_s$ is the Shakura-Sunyaev viscosity parameter, $\dot{M}$ is the mass accretion rate, $M$ is the mass of the accretor, $R$ is 
the radius where the analysis is done, $R_*$ is the radius of the accretor and for a nonrotating black hole it will be the Schwarzschild 
radius ($R_s = 2GM/c^2$; $G$ is gravitational constant, $c$ is the speed of light in free space), \(M_\odot\) is the solar mass. The time 
it takes for a fluid parcel to reach $3R_s$ from 100$R_s$ for a 10\(M_\odot\) accretor is $\sim 8\times 10^3$ seconds, if $\alpha_s = $ 
0.1. The time period of the perturbations will be $(2\pi/\mathcal{R}e(\beta)_{max})\sqrt{R^3/GMq^2}$ seconds. 
From FIG.~\ref{fig:WKB_kep_argand_re_10_4_v_10}, we obtain $\mathcal{R}e(\beta)_{max}$ to be 1.03, 0.48 and 0.3 for $\alpha_1$ to be 1, 5 
and 10 respectively. Time period corresponding to these cases will be 0.57, 1.22 and 1.95 seconds respectively. These time scales are very 
tiny in comparison with the infall time scale of the matter to fall from 100$R_s$, i.e. the fluid parcel gets enough time to be unstable 
before it ultimately falls into the black hole. Our theory, therefore, passes the first check. 

Now the crucial and more important point is 
how much time the fluid parcel takes to cross the shearing box itself along the radial direction. We consider the size of the shearing box 
to be $0.05R_s$ (see \cite{Nath_2015}) and it is situated at $100R_s$. With these configurations, the fluid parcel takes around 5.34 
seconds to cross the box. This time scale is also greater than the time period of the temporal oscillation of the perturbation.

Apart from the timescale corresponding to temporal oscillation of the perturbation, there is another time scale involved in the system and it is at 
which time the system enters into the nonlinear regime. From FIG.~\ref{fig:f_t_squre}, it is clear that 
$(\mathcal{R}e(u))^2$ for $\alpha_1=10$ 
enters into the nonlinear regime for the first time at $t \sim 9$. To make it into second, we have to 
multiply it with a factor 
$\sqrt{R^3/GMq^2}$ which is around 0.09 seconds for the considered system. It, therefore, takes 
around 0.8 seconds for the fluid parcel to enter 
into the nonlinear regime if we consider $\alpha_1 = 10,\ Re=10^4,\ m/u_0 = 10$. 

It is very important to have the wavelength of the 
perturbation inside the box. It, therefore, is necessary to have the maximum wavelength of the perturbation to be equal to the size of the 
box. The wavelength of the perturbation along $x$-direction is $2\pi/\alpha_1$ in dimensionless unit. To make it dimensionful, we have to 
multiply the size of the box ($0.05R_s$) with it. Those $\alpha_1$ which are greater than $2\pi$, therefore, describe the best dynamics of 
the fluid parcel inside the box. This is the reason behind showing the temporal evolution of $(\mathcal{R}e(u))^2$ with corresponding 
$Im(\beta)_{max}$ and $\mathcal{R}e(\beta)_{max}$ for fixed $Re$ and $\alpha_1 = 10\ (> 2\pi)$ for $m/u_0 = 10\ {\rm and}\ 10^2$.

\section{Conclusion}
\label{sec:Conclusion}
Instability and hence turbulence, become inevitable for the fluid parcel inside the shearing box at the small region of the 
accretion disk. This instability is also controlled by $Re$ and the strength of the extra force which is white noise with nonzero mean 
($m$). The presence of noise is very natural. It may arise from small thermal fluctuation present in the systems (see e.g. 
\cite{Nath_2016}). The presence of the noise in the systems can be due to the disturbances of arbitrary origins 
\cite{Farrell_1993}. 
However, in the astrophysical context, particularly in accretion disks, the examples of origin of such force could be: the interaction between the dust grains and fluid parcel in protoplanetary disks (e.g. \cite{Henning_1996}); back reactions of outflow/jet to 
accretion disks; external forcing of the disk, i.e. tidal forcing, shock wave debris, outburst, or internal forcing by nonlinear terms
\cite{Ioannou_2001, razdoburdin_2020}. These forces are also expected to be stochastic in nature. 

Once, the instability and therefore turbulence kick in inside the shearing box, we consider the shearing box repeatedly throughout the 
radial extension of the accretion disk and hence the transport of angular momentum can be interpreted in the Keplerian accretion disk. 
However, for plane Couette flow, there is no requirement of infall. Hence, in presence of noise, it is always expected to lead instability.

\section{Acknowledgement} S.G. acknowledges DST India for INSPIRE fellowship. The authors are thankful to 
the referees for their comments and suggestions, which help present the work better. This work is
partly supported by a fund of Department of Science and Technology (DST-SERB) with research 
Grant No. DSTO/PPH/BMP/1946 (EMR/2017/001226).

\bibliographystyle{ieeetr}
\bibliography{WKB}{}

\appendix

\section{Dispersion relations from Orr-Sommerfeld and Squire equations in the Fourier space}
\label{sec:obtaining_des_rel}
The solutions of equations (\ref{eq:gen_velo_pertb}) and (\ref{eq:gen_vorti_pertb}) in the Fourier space will be 
\begin{eqnarray*}
 \begin{split}
  \tilde{\psi}_{\textbf{k},\omega} = \left(\frac{1}{2\pi}\right)^4 \int_{-\infty}^{\infty}{\psi(x)e^{i(\bm{\alpha}\cdot\textbf{r}-\beta t)} 
e^{-i(\textbf{k}\cdot\textbf{r}-\omega t)}}d^3xdt\\
 = \frac{1}{2\pi}\delta(\alpha_2-k_y) \delta(\alpha_3-k_z) \delta(\beta - \omega) \int_{-\infty}^{\infty}{\psi(x)e^{i(\alpha_1-k_x)x}}dx,
 \end{split}
\end{eqnarray*}
where $\psi$ will be any of $u\ \rm{and}\ \zeta$.

We now integrate equations (\ref{eq:Keplerian_velo_Fourier_space}) and (\ref{eq:Keplerian_vorti_Fourier_space}) with respect to 
$\textbf{k}$ and $\omega$. Each term of equation (\ref{eq:Keplerian_velo_Fourier_space}) after the integration, assuming 
WKB approximation, i.e. neglecting second and higher derivatives of $u$ and $\zeta$, is obtained given below.
\begin{enumerate}
 \item \begin{eqnarray*}
        \begin{split}
         \int_{-\infty}^{\infty}{k_yk^2\frac{\partial \tilde{u}_{\textbf{k},\omega}}{\partial k_x}}d^3kd\omega = 
 \int_{-\infty}^{\infty}{k_y(k_x^2+k_y^2+k_z^2)\frac{\partial \tilde{u}_{\textbf{k},\omega}}{\partial k_x}}d^3kd\omega \\
= \frac{1}{2\pi}\int_{-\infty}^{\infty}{k_y(k_x^2+k_y^2+k_z^2)}\delta(\alpha_2 -k_y)\delta(\alpha_3 -k_z) \Bigl(\frac{\partial}{\partial 
k_x}\\ \int_{-\infty}^{\infty}{dx'u(x')e^{i(\alpha_1-k_x)x'}}\Bigr)dk_x dk_y dk_z \\
=  - 2\alpha_1 \alpha_2 u(0) + 2i\alpha_2u^{\prime}(0),
        \end{split}
       \end{eqnarray*}
  
 \item \begin{eqnarray*}
        \int_{-\infty}^{\infty}{i\omega k^2 \tilde{u}_{\textbf{k},\omega}} d^3k d\omega = i\beta \left(\alpha^2u(0)-2i\alpha_1u'(0)\right),
       \end{eqnarray*}

 \item \begin{eqnarray*}
        \int_{-\infty}^{\infty}{2k_xk_y \tilde{u}_{\textbf{k},\omega}} d^3k d\omega = -2i\alpha_2\left(u'(0)+i\alpha_1 u(0)\right),
       \end{eqnarray*}

 \item \begin{eqnarray*}
        \int_{-\infty}^{\infty}{\frac{k^4}{Re}\tilde{u}_{\textbf{k},\omega}} d^3k d\omega = \frac{\alpha^4}{Re}u(0)-\frac{4}{Re}i\alpha_1 
\alpha^2 u'(0),
       \end{eqnarray*}

 \item \begin{eqnarray*}
        \int_{-\infty}^{\infty}{\frac{2ik_z}{q}\tilde{\zeta}_{\textbf{k},\omega}} d^3k d\omega = \frac{2i\alpha_3}{q}\zeta_0,
       \end{eqnarray*}

 \item \begin{eqnarray*}
        \int_{-\infty}^{\infty}{m_1}\delta(\textbf{k})\delta(\omega) d^3k d\omega = m_1.
       \end{eqnarray*}

\end{enumerate}

We collect all these terms and obtain the first part of equation (\ref{eq:dispersion_rel_orr_som_squ}). Following the same method, we 
also obtain the second part of equation (\ref{eq:dispersion_rel_orr_som_squ}) from equation (\ref{eq:Keplerian_vorti_Fourier_space}).

\end{document}